\documentclass[twocolumn,aps,showpacs,nofootinbib,epsfig]{revtex4}
\usepackage{graphicx}
\usepackage{dcolumn}
\usepackage{bm}
\usepackage{amssymb}
\usepackage{amsmath}
\usepackage{graphicx}
\usepackage{epsfig}
\usepackage[utf8]{inputenc} 
\usepackage{ucs} 
\usepackage{amsmath} 
\usepackage{array}
\usepackage{float}
\usepackage{color}
\usepackage[usenames,dvipsnames]{xcolor}
\usepackage{epstopdf}
\usepackage{natbib}
\usepackage{graphicx}
\usepackage{lineno}
\usepackage{slashed}
\usepackage{color}
\usepackage{caption}
\usepackage{hyperref}
\definecolor{red}{rgb}{1.0,0.0,0.0}
\definecolor{blue}{rgb}{0.0,0.0,1}
\newcommand{\T}{\tilde}
\newcommand{\n}{\mathbf}
\newcommand{\bea}{\begin{eqnarray}}
\newcommand{\eea}{\end{eqnarray}}
\newcommand{\nn}{\nonumber}
\newcommand{\etal}{\textit{et al.}}

\newcommand{\csch}{\text{csch}}

\newcommand{\wc}{\omega_c}
\usepackage{soul}
\begin{document}
%
%
\title{Impact of a topological defect and Rashba spin-orbit interaction on the thermo-magnetic and optical properties of a 2D semiconductor quantum dot with Gaussian confinement}

\author{Jorge David Casta\~no-Yepes$^1$, D. A. Amor-Quiroz$^{1,2}$, C. F. Ramirez-Gutierrez$^{3}$, Edgar A. G\'omez$^4$.}
\address{
  $^1$Instituto de Ciencias
  Nucleares, Universidad Nacional Aut\'onoma de M\'exico, Apartado
  Postal 70-543, M\'exico Distrito Federal 04510,
  Mexico.\\
  $^2$Centre de Physique Théorique, École polytechnique, 91128 Palaiseau, France.\\
  $^3$Posgrado en Ciencia e Ingenier\'ia de Materiales, Centro de F\'isica Aplicada y Tecnolog\'ia Avanzada, Universidad Nacional Aut\'onoma de M\'exico Campus Juriquilla, C.P. 76230, Qro., Mexico.\\
  $^4$Programa de F\'isica, Universidad del Quind\'io, Armenia, Colombia.}
%
\begin{abstract}
In this paper, we examine the effect of introducing a conical disclination on the thermal and optical properties of a two dimensional GaAs quantum dot in the presence of a uniform and constant magnetic field. In particular, our model consists of a single-electron subject to a confining Gaussian potential with a spin-orbit interaction in the Rashba approach. We compute the specific heat and the magnetic susceptibility from the exact solution of the Schr\"odinger equation via the canonical partition function, and it is shown that the peak structure of the Schottky anomaly is linearly displaced as a function of the topological defect. We found that such defect and the Rashba coupling modify the values of the temperature and magnetic field in which the system behaves as a paramagnetic material. Remarkably, the introduction of a conical disclination in the quantum dot relaxes the selection rules for the electronic transitions when an external electromagnetic field is applied. This creates a new set of allowed transitions causing the emergence of semi-suppressed resonances in the absorption coefficient as well as in the refractive index changes which are blue-shifted with respect to the regular transitions for a quantum dot without the defect.\\
\\\\
\end{abstract}
\maketitle
\section{Introduction}
The \textit{quantum dots} (QDs) are considered to be the keystone for 
building solid-state nanodevices with applications in quantum information 
technologies, since it is currently possible to control the 
number of electrons in such mesoscopic systems 
\cite{mesoscopic1, mesoscopic2, mesoscopic3, mesoscopic-self-citation}. 
In particular, spin-related phenomena in QD's has been studied in extension 
in the last decades as they are crucial in the semiconductor technology called spintronics \cite{Wolf}. Among these, spin-orbit (SO) coupling mechanisms in semiconductors provide a basis
for device applications and a source of interesting physics, such as the spin transistor~\cite{Datta}.  The Rashba effect is of particular interest as it provides a SO coupling whose tunability allows SO effects to occur in QDs with few electrons \cite{Governale}. Several studies have been realized around the impact of Rashba-SO interaction (SOI):  In fact, some theoretical studies were carried out on the impact of the Rashba-SOI on the optical properties of a disk-like QD in the presence of an external magnetic field within the framework of the density matrix approach \cite{Hosseinpour}. Interestingly, it was found that the resonance peaks on both the absorption coefficients and refraction index shift to the red with increasing strength of the SOI. Also, it has been shown that the SO effects modify the fluctuations of the conductance of a QD consisting of a GaAs heterostructure when a parallel magnetic field is applied~\cite{Halperin}. An interesting behavior in the magnetization and the susceptibility in a parabolic QD at low magnetic fields have been observed. This fact has been attributed from a theoretical point of view as a consequence of the presence of the Rashba term~\cite{Voskoboynikov}. Similar studies have focused on both Rashba and Dresselhaus spin-orbit coupling mechanisms for explaining the level anticrossing in low-dimensional systems~\cite{Denis,Destefani}. In addition, other related works in this field have also investigated the influence of the SOI on the energy levels of electrons within parabolic confinement~\cite{Marian,Voskoboynikov2}. More recently, it has been found that, there is a significant dependence of the Rashba contribution on the electronic, thermo-magnetic and transport properties~\cite{Sanjeev}.

On the other hand, the shape of the electron confining potential is crucial for the correct description of QD dynamics. It is well established that a harmonic 
potential is a good approximation which reproduces the main characteristics of such systems. Recently, Castaño \etal~\cite{Castano} studied the thermal and magnetic properties 
of a parabolic GaAs-QD in the presence of external magnetic and electric fields, resulting in a good description of the thermo-magnetic phase diagram. Kumar~\cite{Sanjeev2} 
included a SOI term and electron-electron interactions to the parabolic potential model (PPM) 
in order to compute the ground state of the GaAs QD. Further experimental investigations have shown that the confining potential is rather anharmonic and has a finite depth which has been simulated by several authors using a Gaussian potential model (GPM)~\cite{Adamowski,GaussianPotential1,GaussianPotential2,GaussianPotential3,GaussianPotential4,GaussianPotential5,GaussianPotential6}. In fact, this potential model has been widely used in several branches of physics, for example, in the description of the Gaussian core model of interacting particles~\cite{Louis:2000,Prestipino:2005}, and stability diagrams in double quantum dot systems~\cite{Wang:2011}.

Furthermore, several studies have been carried on the optical properties of systems composed of a two dimensional QD (2D-QD). For instance, the effect of different geometries, such as quantum rings \cite{anillos1, anillos2} and triangular QDs \cite{triangular_QD} have proved to have a direct impact on the relevance of nonlinear optical effects. The optical absorption coefficients  in a disk-like QD have been shown to be 2 to 3 orders of magnitude higher than those of a spherical QD \cite{spherical_QD}. In recent years there has been growing interest on the effects of the non-resonant intense laser field on the optical properties in semiconductor QDs~\cite{sari:2017}, as well as the influence of the electric and magnetic fields in the second (third) harmonic generation~\cite{Yesilgul:2017,Sakiroglu:2018,Orozco:2017,Restrepo:2017}. More recently, theoretical studies have allowed characterizing the conduction band states of an electron in elliptically shaped quantum rings with possible applications in electronics and optoelectronics~\cite{Vinasco:2018,Vinasco2:2018}. 
On the other hand, the presence of topological defects in a material have been gaining much attention from a theoretical point of view, since they can modify the electrical, acoustic or thermal
properties in the material~\cite{esquemacono}. For example, the presence of such defects in graphene nanoribbons have shown a significant reduction in the thermal conductivity~\cite{Haskin:2011}, as well as the possibility to manipulate nanoparticles in arrays of topological defects~\cite{Kasyanyuk:2016}. In the context of the condensed matter physics, the topological defects have also been used for studying a variety of quantum systems, ranging from superfluid helium~\cite{Alexander:2012} and its plentiful spontaneous symmetry breakings~\cite{Volovik:2003} to liquid crystals~\cite{Pereira2:2013} or mesoscopic physics~\cite{Sinha:2016}.

The aim of this work is to investigate the thermal, magnetic and optical properties of a single-electron system incorporating the Rashba SOI within a 2D-QD that displays a topological defect given by a conical disclination~\cite{Deser, Mishra,Rojas2018}. Moreover, our model consists of a particle trapped in a potential that interpolates between  parabolic and a Gaussian potential, as the whole system is subjected to a uniform external magnetic field. This paper is organized as follows. In Sec.~\ref{sec:formalism} we present the theoretical model for describing a single-electron subject to a confining Gaussian potential with a spin-orbit interaction in the Rashba approach. Additionally, a topological defect is included in the model by considering a conical disclination. In the same section, we compute both thermal and magnetic properties via the canonical partition function as well as the optical properties related to the total absorption coefficient and the refractive index changes. In Sec.~\ref{Results_discussion}, we present and discuss our numerical results which are summarized in Sec.~\ref{sec:conclusions}.
\section{Theoretical model}\label{sec:formalism}
\subsection{Energy spectrum of the quantum system}
\begin{figure}[ht!]
\centering\includegraphics[scale=0.3]{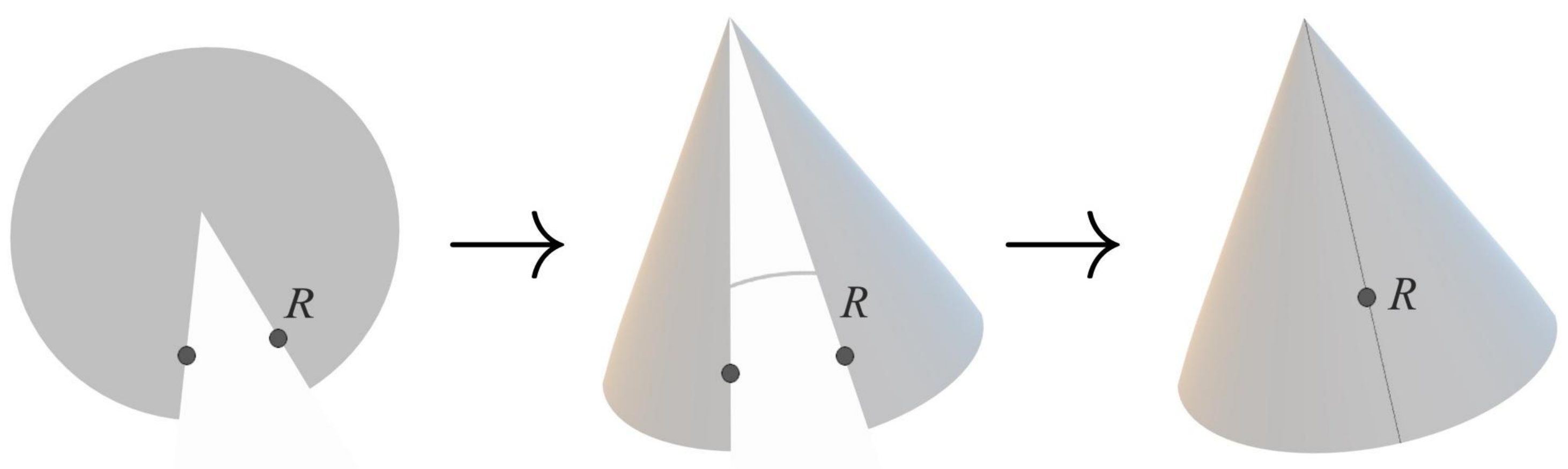}\caption{Volterra construction. A cut planar disk with two radial lines identified point-to-point (marked as bold dots) is isometric to a cone. The evaluated system remains 2-dimensional though the angular periodicity of the wave function is recovered in the $(r,\theta)$-coordinate system.}
\label{Volterra}
\end{figure}

In the present model, we study a single-electron 2D-QD in the presence of an external uniform magnetic field. The surface has a topological defect described in polar coordinates by the metric
\bea
dl^2=d\rho^2+\rho^2d\phi^2, \;\;\text{with}\;\;\phi\in[0,2\pi\alpha].
\label{metric1}
\eea
The parameter $\alpha\in(0,1)$ controls the cut-off and thus 
$\alpha=1$ would reduce to the case with no topological defect.
As it also quantifies the conicity of the surface it is known 
as the kink parameter. 
\begin{figure}[ht!]
\centering
\includegraphics[scale=.4]{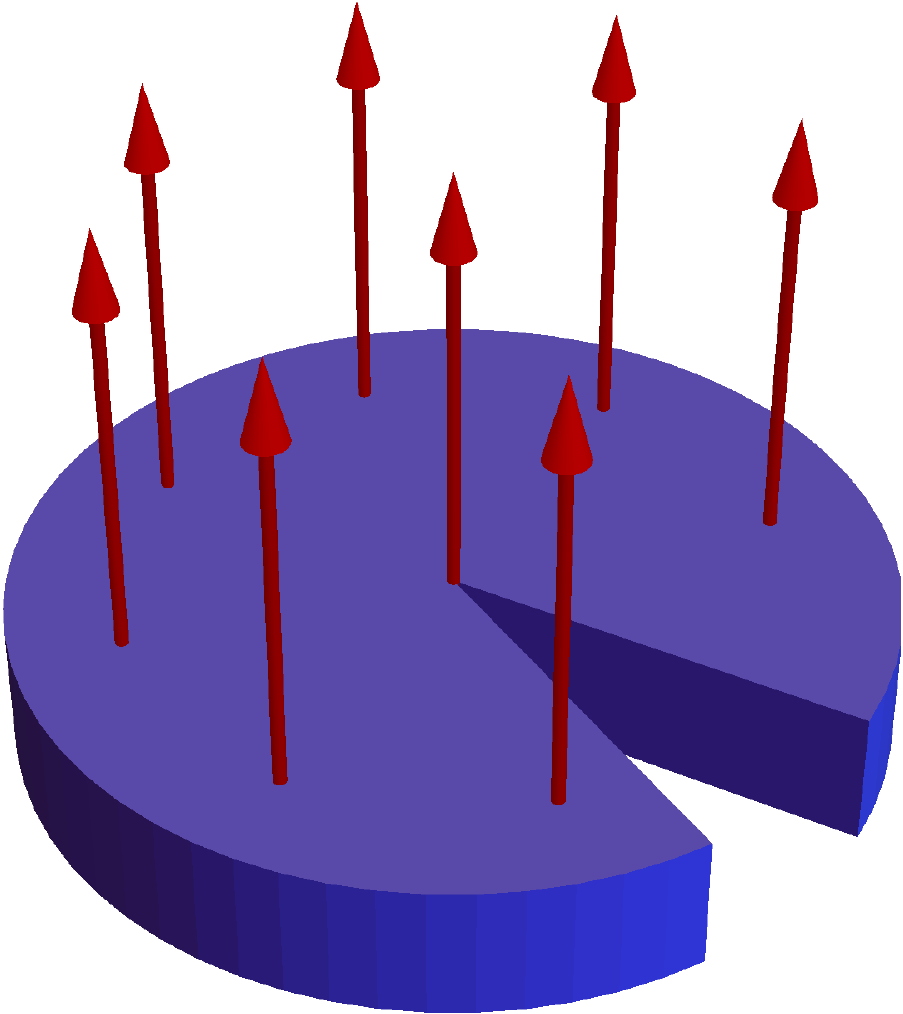}
\caption{A conical disclination is constructed by slicing off a sector with a certain apex angle out off a flat space. Said angle is referred in the literature as \textit{deficit angle}. The red arrows represent the direction of the homogeneous magnetic field, which is chosen to be normal to the surface. }\label{esquema}
\end{figure}
The Fig.~\ref{esquema} schematizes the system to be studied. The angular limitation leading to an excised region complicates the analysis, though a transformation to a new coordinate system can be 
achieved via the set of transformation equations~\cite{Deser, Mishra,Rojas2018}
\bea
r = \alpha \rho,
~~~~
\theta = \alpha^{-1}\phi ,
\label{cambiodecoords}
\eea
where $\theta\in[0,2\pi]$. Therefore the metric becomes
\bea
dl^2=\alpha^{-2}dr^2+r^2d\theta^2, \;\;\text{with}\;\;\theta\in[0,2\pi] .
\label{metric2}
\eea
The Hamiltonian of the system in the presence of an external magnetic field with both Zeeman and Rashba SO terms is given by
\bea
\hat{H}&=&\frac{1}{2m^*}\left(\n{p}-\frac{q}{c}\n{A}\right)^2+\hat{H}_{\text{Gauss}}\nn\\
&+&\hat{V}_{SOI}(\rho,\phi)+\frac{1}{2}\mu_Bg^*\hat{\n{S}}\cdot \n{B},
\eea
where $q$ and $c$
denote the charge of the electron and the speed of light in vacuum, respectively. Moreover, $m^*$ is the effective electron mass which is taken as a constant in order to neglect the non-parabolicity of the conduction band. Consequently, the effects produced by the conduction electrons are not taken into account in the present work~\cite{conductionband1,conductionband2,conductionband3}. The term $\hat{H}_{\text{Gauss}}$ corresponds to the confining potential, $\hat{V}_{SOI}(\rho,\phi)$ is the SOI and the last term is the Zeeman coupling of the external magnetic field ${\mathbf{B}}$ with the electron spin ${\mathbf{\hat{S}}}$. Here $\mu_B$ is the Bohr magneton and $g^*$ is the effective Land\'e factor of the electron. The vector potential $\hat{A}$ is expressed in the symmetric gauge $\hat{A}=\frac{B}{2}(-y,x,0)$ which in the $(r,\theta)$-coordinate system has the form
\bea
\n{A}(r)=\frac{Br}{2\alpha}\hat{e}_{\theta}.
\eea
The confining potential $\hat{H}_{\text{Gauss}}$
consists of a parametrization that interpolates between a parabolic and a Gaussian potential as shown in recent theoretical works~\cite{GaussianPotential1,GaussianPotential2}. In such manner, the Gaussian model can be approximated as a parabolic potential plus a perturbation:
\bea
\hat{H}_{\text{Gauss}}&=&-V_0 e^{-\rho^2/2R^2},\nn\\
&\approx&\frac{m^*}{2}\left[(1-\kappa)\omega_h^2+2V_0\kappa \left(\frac{\Tilde{\omega}}{\hbar+2m^*\Tilde{\omega}R}\right)\right]\rho^2\nn\\
&-&V_0,\nn\\
&=&\frac{m^*}{2\alpha^2}\,\omega^2r^2-V_0.
\eea
where $V_0$ and $R$ define the depth and the range of the potential, respectively. Notice that these two physical parameters define the effective
size of the semiconductor QD. In particular, we have defined
\bea
\omega^2=(1-\kappa)\omega_h^2+2V_0\kappa \left(\frac{\Tilde{\omega}}{\hbar+2m^*\Tilde{\omega}R}\right)
\eea 
with $\hbar$ the Planck constant. Additionally, $\Tilde{\omega}=\sqrt{\omega_c^2+\omega_h^2}$, $\omega_c=qB/m^*$, and $\omega_h^2=V_0/m^* R^2$. Here the parameter $\kappa$ controls the form of the confinement: $\kappa=0$ for the PPM and $\kappa=1$ for the GPM. The SOI term is given by the general expression~\cite{SOIBerry,SO1}
\bea
\hat{V}_{SOI}&=&\frac{\gamma_s}{\hbar}\vec{\sigma}\cdot\left[\nabla V\times\left(\n{p}-\frac{q}{c}\n{A}\right)\right],
\eea
where the normal to the surface is chosen along the $z$-axis. The Rashba spin-orbit coupling is denoted by $\gamma_s$ and $\vec{\sigma}=(\sigma_x,\sigma_y,\sigma_z)$ is the Pauli matrices vector. Thus, $\hat{V}_{SOI}$ in the coordinate representation has the form
\bea
\hat{V}_{SOI}(\rho,\phi)&=&\gamma_s\sigma_z\frac{dV_c}{d\rho}\left[-i\left(\frac{1}{\rho}\right)\frac{\partial}{\partial\phi}+\frac{q}{2\hbar}B\rho\right].
\eea
The confining potential $V_c$ corresponds to the Gaussian potential $\hat{H}_{\text{Gauss}}$. Therefore, we have that
\bea
\hat{V}_{SOI}(r,\theta)&=&\frac{m^*}{2}s\gamma_s\frac{m^*\,\omega_c\, \omega^2}{\alpha^2\hbar}r^2\nn\\
&-&i\sigma_z\frac{\gamma_s m^*\,\omega^2}{\alpha}\frac{\partial}{\partial\theta},
\eea
with $s=\pm1$ referring to the spin projection. Finally, the Hamiltonian of the system becomes
\bea
\hat{H}&=&-\frac{\hbar^2}{2m^*}\left[\frac{\alpha^2}{r}\frac{\partial}{\partial r}\left(r\frac{\partial}{\partial r}\right)+\frac{1}{r^2}\frac{\partial^2}{\partial\theta^2}\right]+\frac{\wc}{2\alpha^2}\hat{L}_z\nn\\
&+&\frac{1}{2}\frac{m^*}{\alpha^2}\Omega^2_s (\alpha,\kappa)\,r^2-V_0\nn\\
&+&\frac{1}{2}\mu_Bg^*\hat{\n{S}}\cdot \n{B}-i\sigma_z\frac{\gamma_s m^*\,\omega^2 }{\alpha}\frac{\partial}{\partial\theta}.
\label{Hamiltonian}
\eea
The effective harmonic frequency in Eq.~(\ref{Hamiltonian}) is given by
\bea
\Omega^2_s=
\Omega^2_s(\alpha,\kappa)=\left(1+s\gamma_s\frac{m^*\wc}{\hbar}\right)\omega^2+\left(\frac{\omega_c}{2\alpha}\right)^2.
\label{Omega_S}
\eea
It is worth mentioning that the axial symmetry of this problem remains unaltered even when the angular defect is considered. 
This fact can be well understood within the context of the differential geometry and topological defects, where a cut planar disk with two radial lines (denoted by 
$R$ in Fig.~\ref{Volterra}) is isometric to a cone through Volterra cut-and-glue constructions~\cite{Volterra2,Sinha:2016,esquemacono,Volterra1}. This symmetry appears naturally in the Hamiltonian given by  Eq.~(\ref{Hamiltonian}), where there is no explicit dependence on the angle. It commutes with the generator of rotations in the $(r,\theta)$-space and therefore, we admit a periodic solution in the $\theta$-coordinate as follows:
%
\bea
\psi(r,\theta)=\frac{1}{\sqrt{2\pi}}e^{il\theta}R_{nls}(r)\chi_s(\sigma), \;\;\text{such that}\;\;l\in\mathbb{Z}. 
\eea
%
With the above, the eigenvalues equation is
\begin{widetext}
\bea
-\frac{\hbar^2}{2m^*}\left[\frac{\alpha^2}{r}\frac{d}{d r}\left(r\frac{d}{d r}\right)-\frac{l^2}{r^2}\right]R_{nls}(r)+\left[\frac{m^*\Omega^2}{2\alpha^2}r^2-V_0+\frac{1}{2}\frac{\hbar\wc}{\alpha^2}l+\left(\gamma_s m^*\omega^2l+\frac{1}{4}g^*\hbar\omega_c\right)s\right]R_{nls}(r)=E_{nls}R_{nls}(r). \quad
\label{radialec}
\eea
\end{widetext}

The exact solutions for the eigenfunctions and eigenvalues are:
\begin{subequations}
\bea
R_{nls}(\T{r})=\left(\frac{m^*\Omega_s}{\hbar}\right)^{1/2}\sqrt{\frac{2n!}{\left(n+p|l|\right)!}}e^{-\frac{\T{r}^2}{2}}\T{r}^{p|l|}L_n^{p|l|}\left(\T{r}^2\right),\nn\\
\eea
\bea
E_{nls}&=&\hbar\Omega_s \left(2n+p|l|+1\right)+\frac{p^2}{2}\hbar\wc l\nn\\
&+&\left(\gamma_s m^*\omega^2l+\frac{1}{4}g^*\hbar\omega_c\right)s-V_0,
\label{energyspectrum}
\eea
\end{subequations}
where $\T{r}=\sqrt{m^*\Omega_s/\hbar}\;r$ and $p=\alpha^{-1}$ is the inverse kink parameter.
\subsection{Thermal and magnetic properties}
The canonical partition function is calculated from the energy spectrum in Eq.~(\ref{energyspectrum}):
\bea
\mathcal{Z}&=&\sum_{n=0}^{+\infty}\sum_{l=-\infty}
^{+\infty}\sum_{s=-1}^1 e^{-\beta E_{nls}},\nn\\
&=&\mathcal{Z}^{(+)}+\mathcal{Z}^{(-)},
\eea
where $\beta=1/k_BT$  is the inverse temperature and $k_B$ the Boltzmann’s constant. The sum can be analytically performed and results in the following expression
\begin{subequations}
\bea
\mathcal{Z}^{(\pm)}&=&\frac{1}{4}\exp\left[\beta\left( V_0\mp\frac{1}{4}g^*\hbar\wc\right)\right]\sinh\left[p\beta\hbar\Omega_{\pm1}\right]\nn\\
&\times&\csch\left[\beta\hbar\Omega_{\pm1}\right]\csch\left\{\frac{\beta}{2}\left[b^{(\pm)}-p\hbar\Omega_{\pm1}\right]\right\}\nn\\
&\times&\csch\left\{\frac{\beta}{2}\left[b^{(\pm)}+p\hbar\Omega_{\pm1}\right]\right\},\nn\\
\eea
with $\Omega_{\pm 1}$ as defined in Eq.~(\ref{Omega_S}) and the shorthand notation
\bea
b^{(\pm)}=\frac{p^2}{2}\hbar\wc\pm\gamma_s m^*\omega^2.
\eea
\end{subequations}
In the present work, we focus in the specific heat and the magnetization of the system which can be easily obtained by 
\bea
C_v=k_B\beta^2\frac{\partial^2}{\partial\beta^2}\ln\mathcal{Z}\;\;\text{and}\;\;\chi=\frac{1}{\beta}\frac{\partial^2}{\partial B^2}\ln\mathcal{Z},
\label{funcionestermodinamicas}
\eea
respectively. As it is required for the discussion in Section~\ref{Results_discussion}, we define the Schottky  temperature $T_s$ as
\bea
\frac{\partial C_v}{\partial T}\bigg|_{T_s}=0
, ~~~ \text{and} ~~~
\frac{\partial^{2} C_v}{\partial T^2}\bigg|_{T_s} < 0 .
\label{Schotkky_temp}
\eea
Such temperature corresponds to the low-energy peak present in the Schottky anomaly. It is closely related to the energy required for a thermal transition between the ground and the first excited state (with energy $\Delta E$) as it can be interpreted as a resonance in $k_B T_s \sim \Delta E$.
\subsection{Optical Properties}
\begin{figure}[h!]
\centering
\includegraphics[scale=0.39]{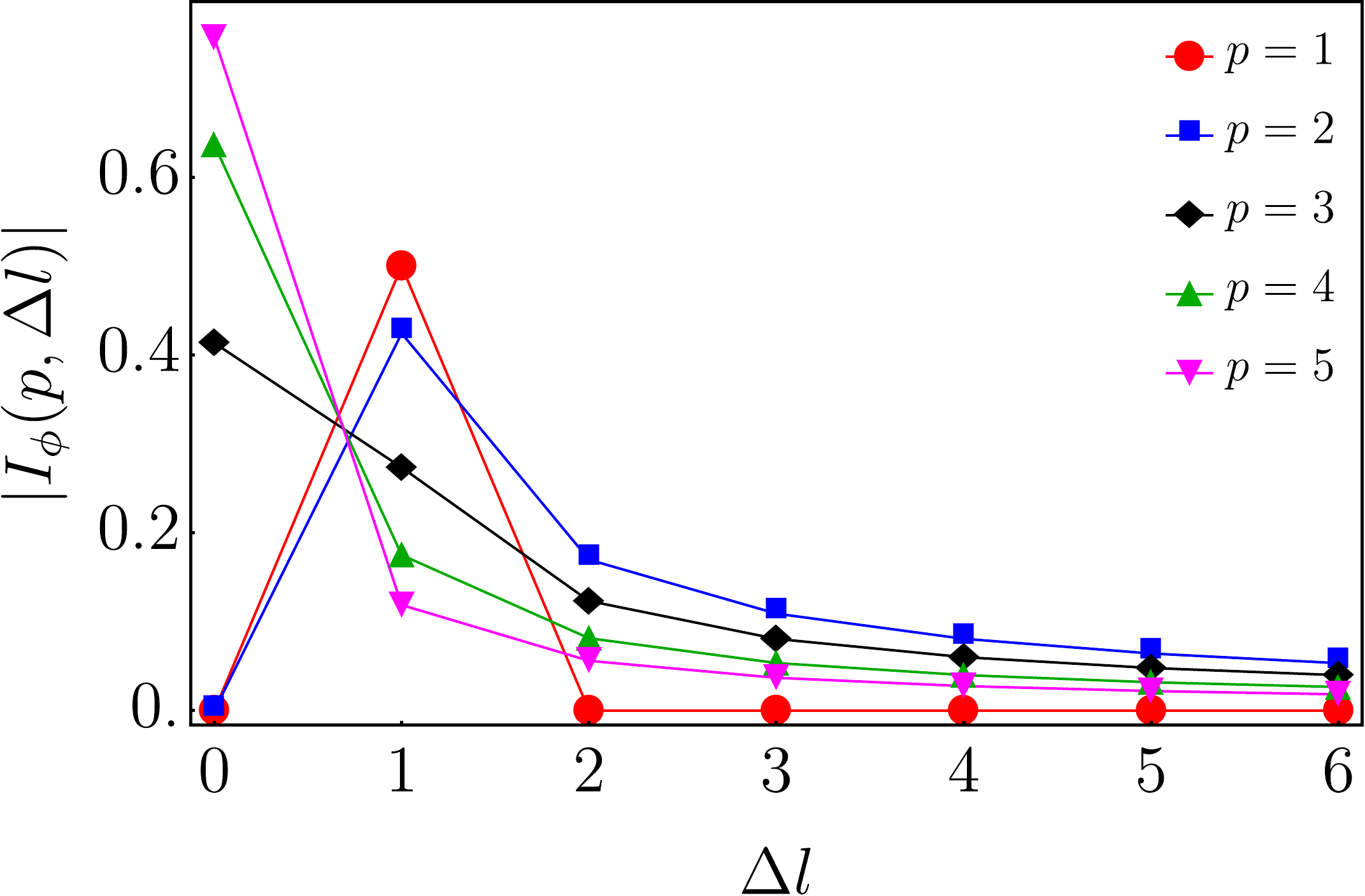}
\caption{Variation of the angular integral $I_\phi(p,\Delta l)$ defined in Eq.~(\ref{transition_def}) for different values of the inverse kink parameter $p$. One can notice that the transitions are in general allowed even for large values of $\Delta l$, though they decrease in probability after a certain value of $\Delta l$.}
\label{Iphi}
\end{figure}
\begin{figure*}[ht!]
\centering
\includegraphics[scale=0.38]{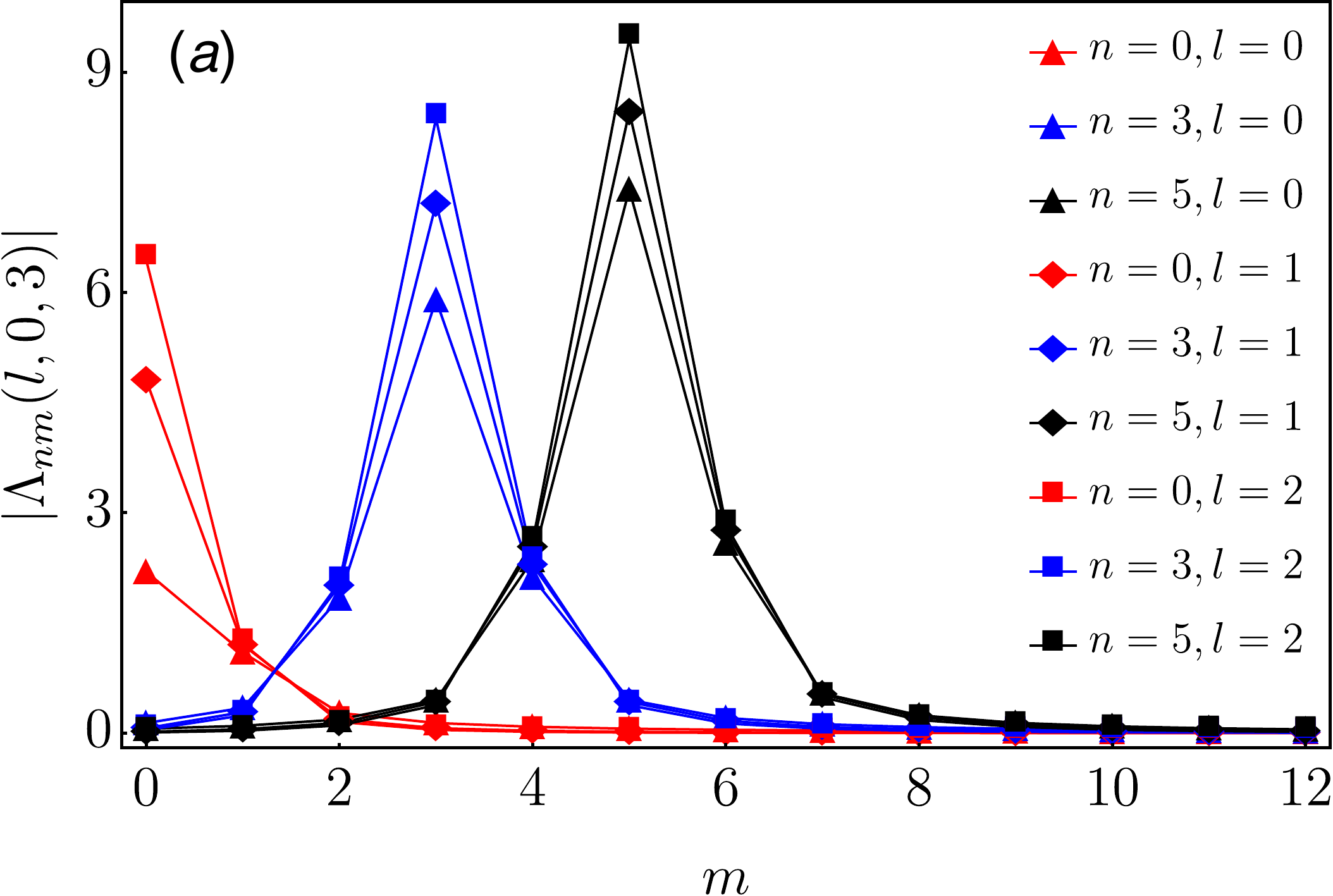}\;\;\;\;\includegraphics[scale=0.38]{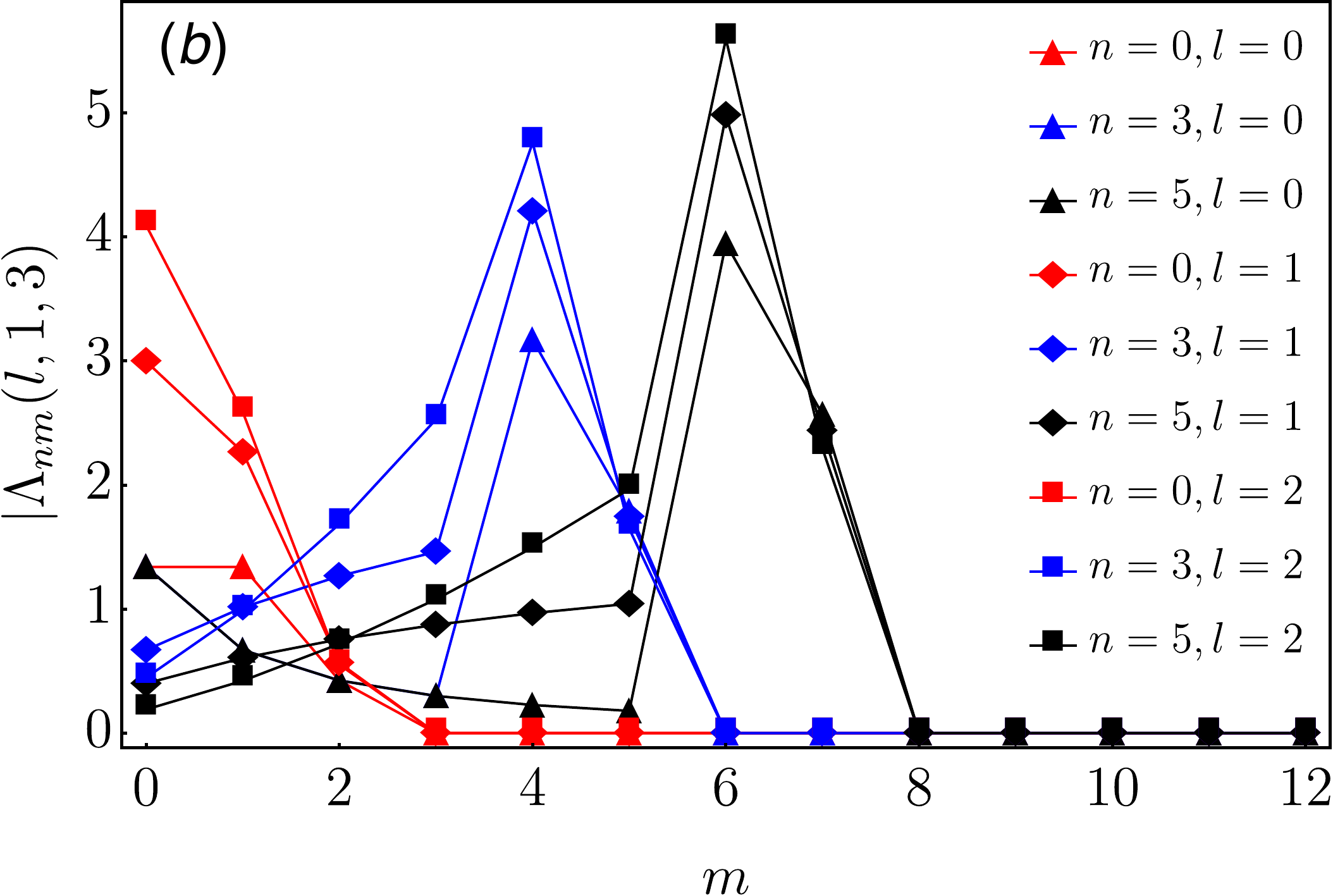}
\caption{Reduced dipole matrix elements $\Lambda_{nm}(l,\Delta l, p)$ as defined in Eq. (\ref{transition_def}), where $n$ and $m$ refer to the radial quantum number, while we denote the angular momentum difference as $\Delta l=l'-l$. Panel (a) shows the numerical calculation for $p=3$ and $\Delta l =0$, and panel (b) shows the case for $p=3$ and $\Delta l =1$. The different values of $n$ and $l$ as labelled in the legend.}
\label{LambdaAB}
\end{figure*}
The refractive index changes and optical absorption coefficient of the 2D-QD are calculated by using the density matrix formalism iteration approach~\cite{Opticalproperties1,Opticalproperties2, Opticalproperties3}. We assume that the system is excited by an external electromagnetic plane wave of frequency $\omega_0$ and polarized in the $\n{\hat{x}}$ direction
\bea
{\bf E}(t)={\bf E}_0\cos(\omega_0 t)=\Tilde{\bf{E}}e^{i\omega_0 t}+\Tilde{\bf{E}}^*e^{-i\omega_0 t}.
\eea
The linear and the third-order nonlinear refractive index changes can be expressed as %
\bea
\frac{\Delta n^{(1)}(\omega_0)}{n_r}&=&\frac{\sigma_v\left|M_{ji}\right|^2}{2n_r^2\epsilon_0}\left[\frac{E_{ji}-\hbar\omega_0}{(E_{ji}-\hbar\omega_0)^2+\left(\hbar\Gamma_{ji}\right)^2}\right],\nn\\
\frac{\Delta n^{(3)}(\omega_0)}{n_r}&=&-\frac{\sigma_v\left|M_{ji}\right|^2}{4n_r^3\epsilon_0}\frac{\mu c I}{\left[(E_{ji}-\hbar\omega_0)^2+\left(\hbar\Gamma_{ji}\right)^2\right]^2}\nn\\
&\times&\Bigg{\{}4(E_{ji}-\hbar\omega_0)\left|M_{ji}\right|^2-\frac{\left(M_{jj}-M_{ii}\right)^2}{\left(E_{ji}\right)^2+\left(\hbar\Gamma_{ji}\right)^2}\nn\\
&\times&\Bigg{[}\left(E_{ji}-\hbar\omega_0\right)\left[E_{ji}\left(E_{ji}-\hbar\omega_0\right)-\left(\hbar\Gamma_{ji}\right)^2\right]\nn\\
&-&\left(\hbar\Gamma_{ji}\right)^2\left(2E_{ji}-\hbar\omega_0\right)\Bigg{]}\Bigg{\}},
\eea
where $\mu$ is the permeability of the system defined as $\mu=1/\epsilon_0 c^2$, with $\epsilon_0$ the electrical permittivity of the vacuum. Additionally, $\sigma_v$
is the carrier density and the incident optical intensity $I$ defined as $I=2\sqrt{\epsilon_r/\mu}\vert \tilde{E}(\omega_0) \vert ^2$. Notice that $\epsilon_r$ is the real part of the permittivity which is defined through $\epsilon_r=n^2_r\epsilon_0$ with $n_r$ the medium refractive index. The term
$M_{ji}=\langle j|qx|i\rangle$ represents the electric dipole moment matrix element, $E_{ji}=E_j-E_i$ is the energy difference between $i$-th and $j$-th electronic levels, and $\hbar \omega_0$ the incident photon energy.
\\\\
The total refractive index change can be written as
\begin{figure*}[t]
\centering
\includegraphics[scale=0.39]{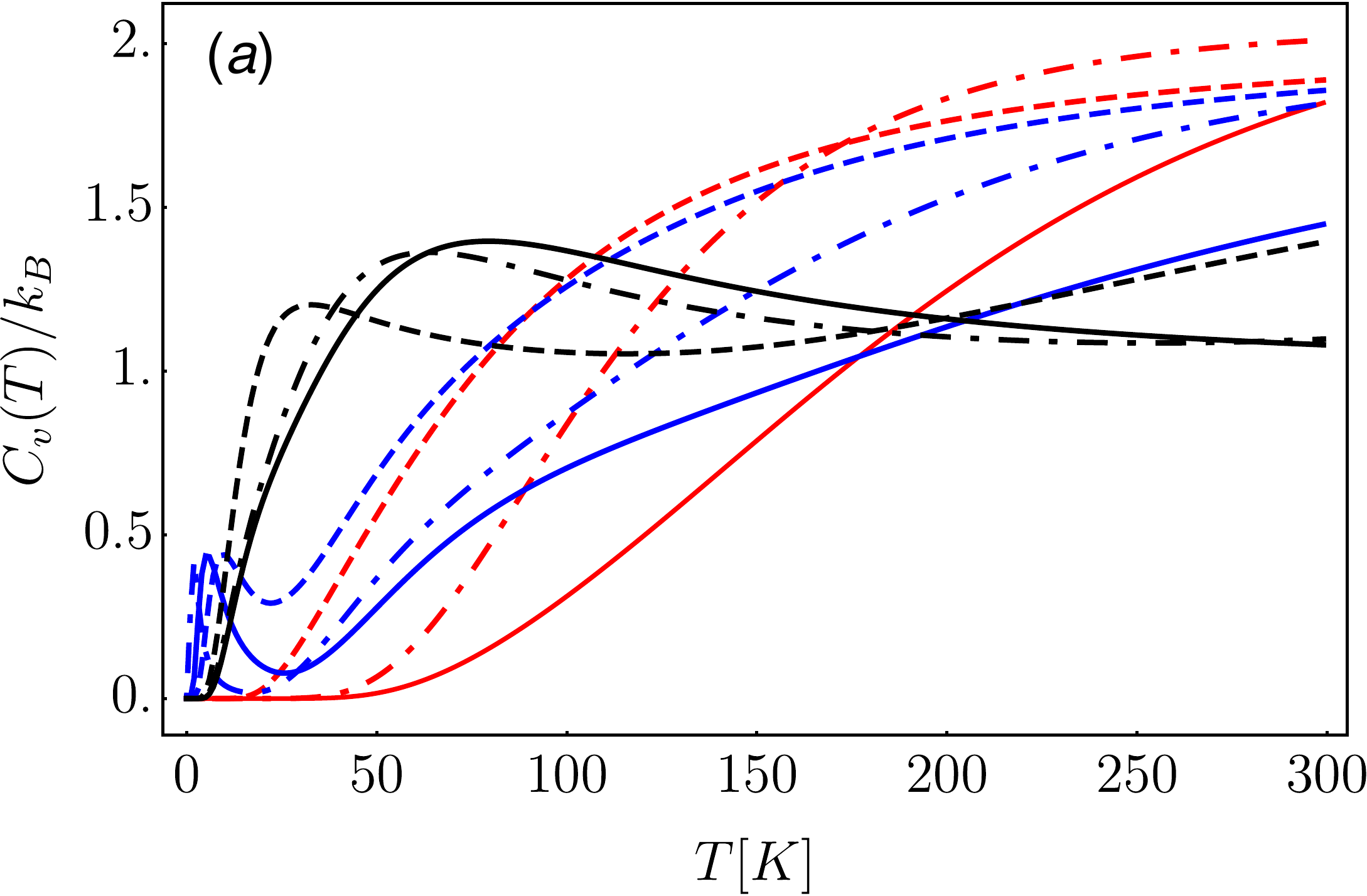}\;\;\;\;\includegraphics[scale=0.39]{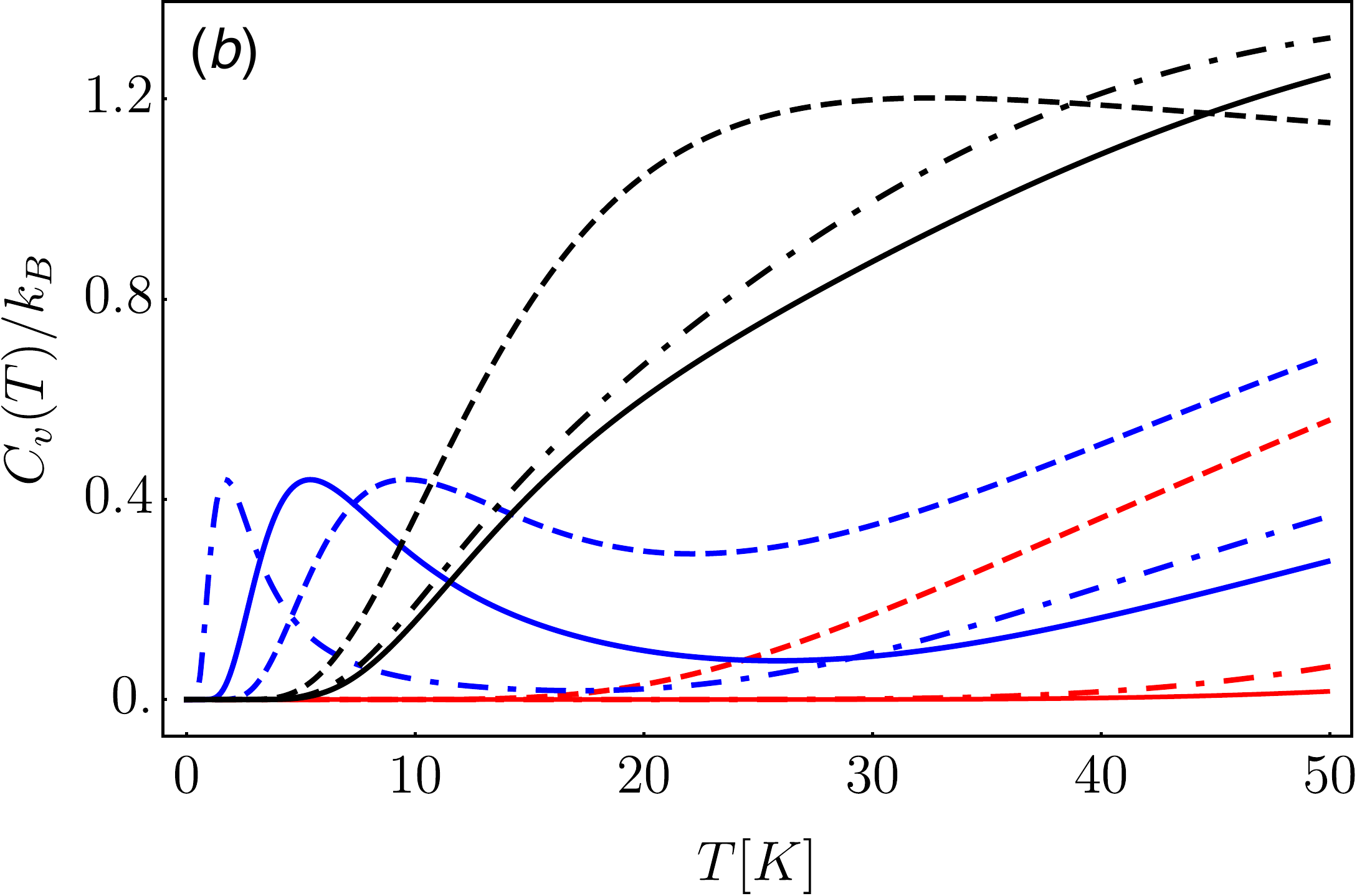}\caption{
Panel (a) shows the specific heat $C_v$ as a function of the temperature with the parameters
$p=4$, $\kappa=1$ and $\gamma_s=20$ nm$^2$ for $B=0$T (solid red line), $B=10$T (solid blue line) and $B=50$T (solid black line), respectively. Numerical results obtained with the parameters $p=1$, $\kappa=0$ and $\gamma_s=0$ nm$^2$ for $B=0$T (dashed red line), $B=10$T (dashed blue line) and $B=50$T (dashed black line), respectively. Numerical results obtained with the parameters $p=2$, $\kappa=1$, $\gamma_s=20$ nm$^2$ for $B=0$T (dot-dashed red line), $B=10$T (dot-dashed blue line) and $B=50$T (dot-dashed black line), respectively. Panel (b) displays the details in the low temperature regime where the Schottky anomaly is discernible.  Color coding as in panel (a).}
\label{CvvsT}
\end{figure*}
%
\bea
\frac{\Delta n(\omega_0)}{n_r}=\frac{\Delta n^{(1)}(\omega_0)}{n_r}+\frac{\Delta n^{(3)}(\omega_0)}{n_r}.
\eea
In the same order of approximation, the linear and third-order nonlinear absorption coefficients are
\bea
\alpha^{(1)}(\omega_0)&=&\omega_0\sqrt{\frac{\mu}{\epsilon_r}}\left [ \frac{\sigma_v \hbar \Gamma_{ji}\left | M_{ji} \right |^2}{\left ( E_{ji}-\hbar \omega_0 \right )^2+ \left( \hbar  \Gamma_{ji} \right )^2} \right],\nn\\
\alpha^{(3)}(\omega_0,I)&=&-\omega_0\sqrt{\frac{\mu}{\epsilon_r}}\left ( \frac{I}{2\epsilon_0 n_r c} \right )\nn\\
&\times& \frac{\sigma_v\hbar\Gamma_{ji}\left | M_{ji} \right |^2}{\left [ \left ( E_{ji}-\hbar \omega_0 \right )^2+ \left( \hbar  \Gamma_{ji} \right )^2 \right ]^2}\nn \\
&\times& \Bigg{\{}\frac{\left | M_{jj}-M_{ii} \right |^2\left [4E_{ji}\hbar \omega_0 -\hbar^2\left ( \omega_0^2-\Gamma_{ji}^2 \right ) \right ]}{E_{ji}^2+(\hbar\Gamma_{ji})^2}\nn\\
&-&\frac{3E_{ji}^2\left | M_{jj}-M_{ii} \right |^2 }{E_{ji}^2+(\hbar\Gamma_{ji})^2}+ 4\left | M_{ji} \right |^2\Bigg{\}},\nn\\
\eea
where $\Gamma_{ji}=1/\tau_{ji}$ is the relaxation rate defined through the relaxation time $\tau_{ji}$. A well-known fact in semiconductor nanostructures is that the relaxation rate is strongly related not only to the materials constituting the QDs, but also to some other factors, such as the boundary conditions, temperature of the system and collision processes associated to  electron--impurity, electron--phonon and  electron--electron interactions.
\begin{figure*}[t]
\centering
\includegraphics[scale=0.45]{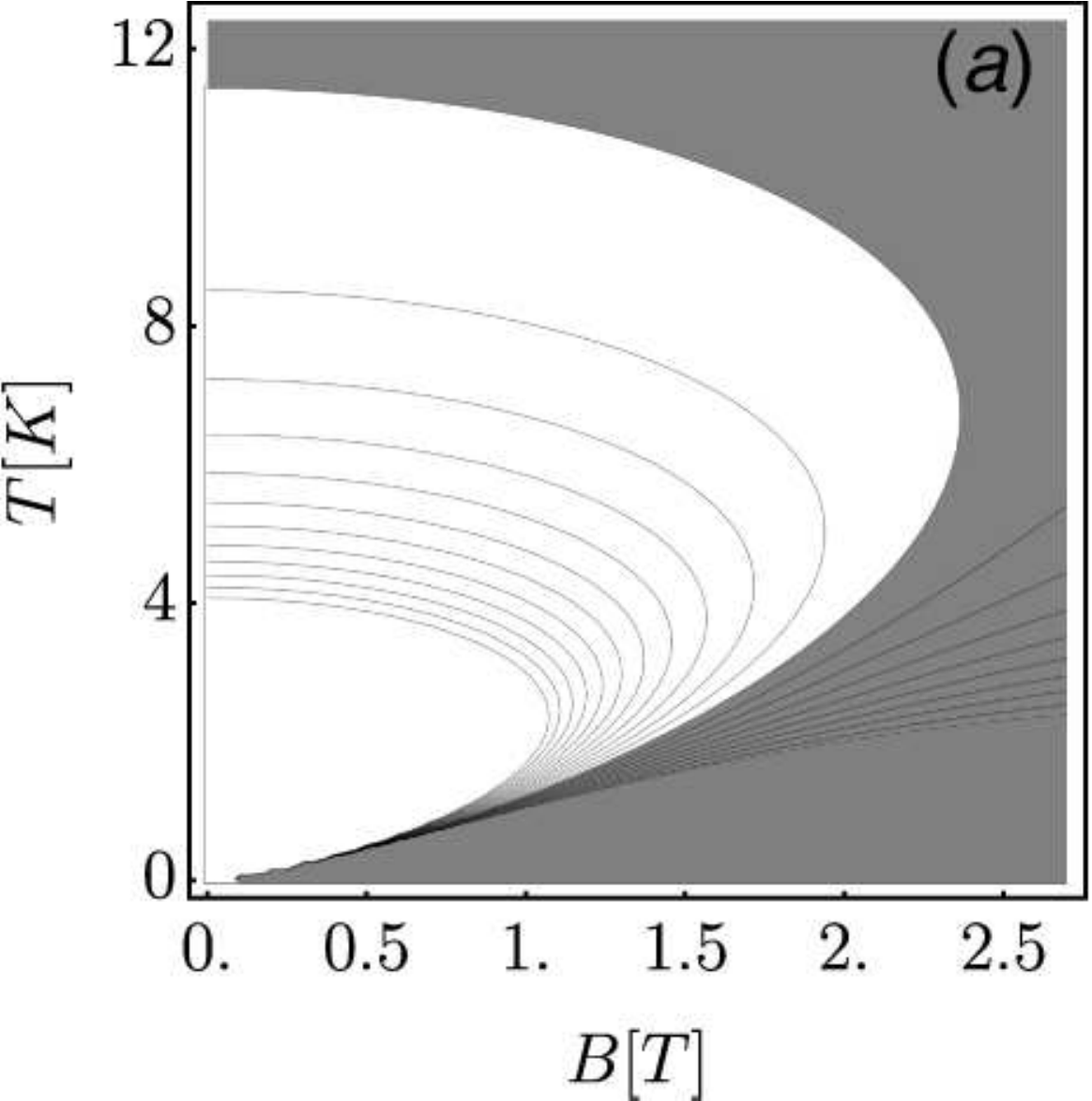}\;\;\;\;\;\;\includegraphics[scale=0.46]{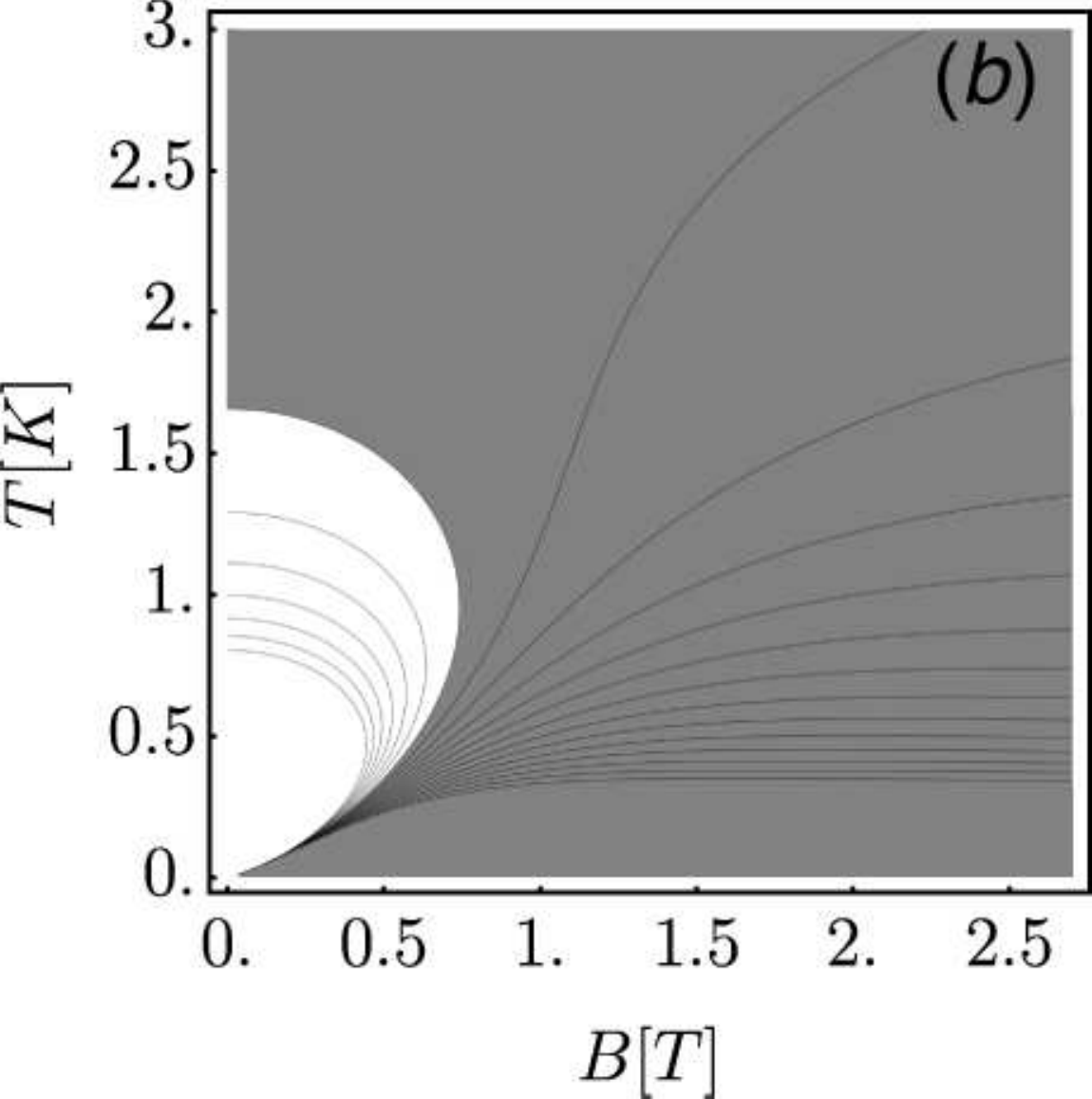}\\
\includegraphics[scale=0.46]{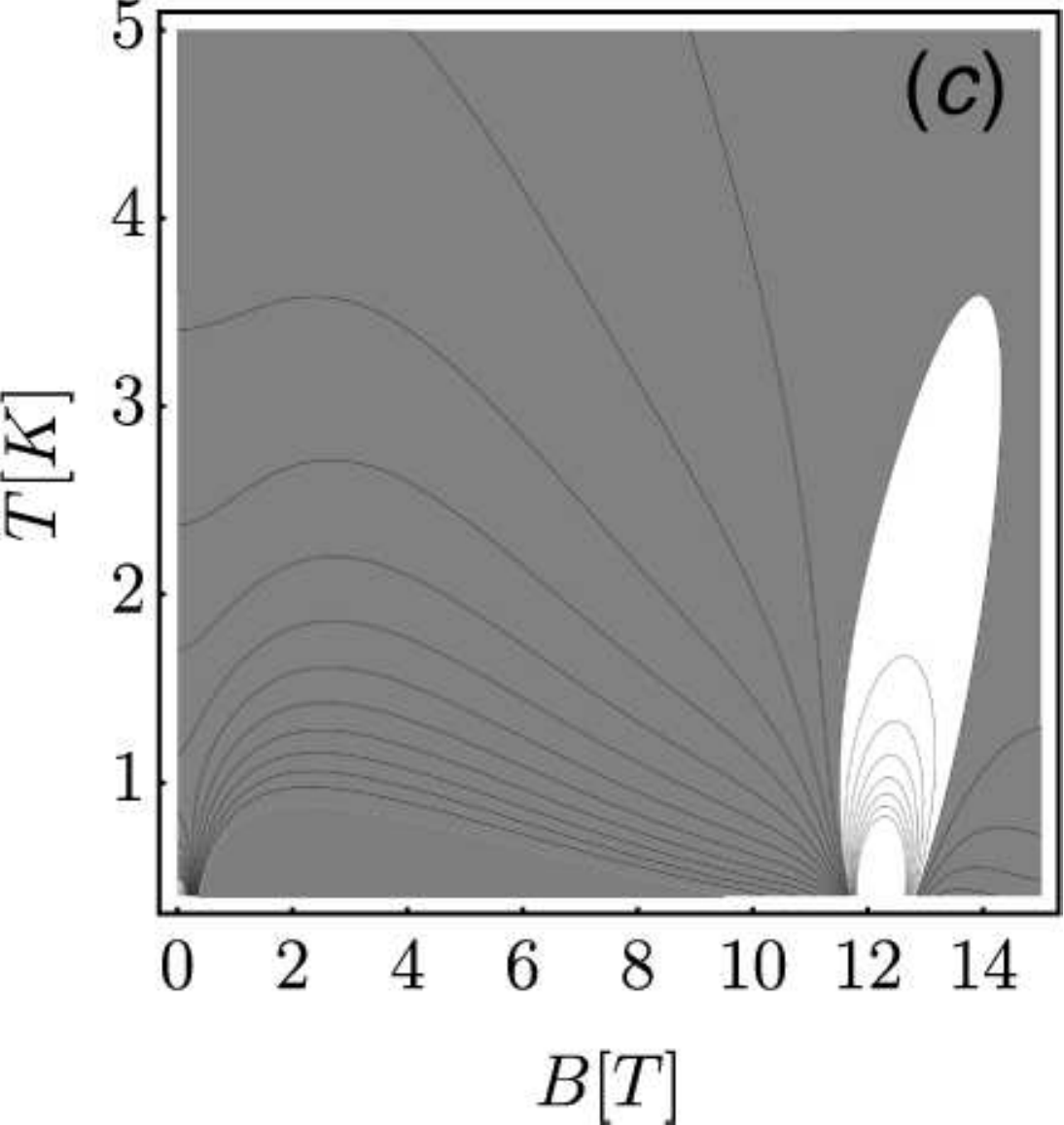}\;\;\;\;\;\;\includegraphics[scale=0.46]{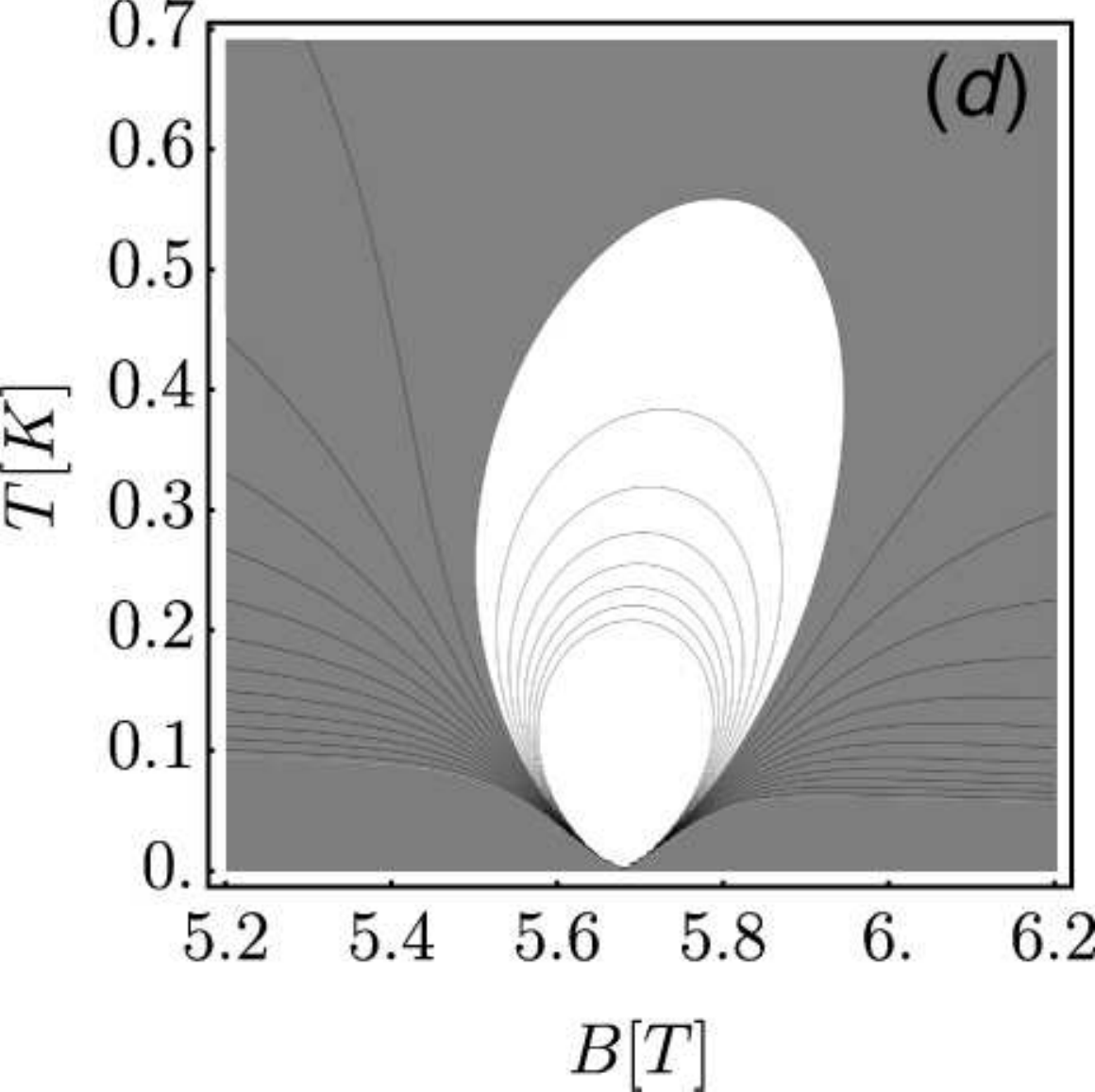}
\caption{Magnetic phase diagram of GaAs quantum dot as function of the temperature and the magnetic field. In panel (a) the parameters are fixed to $\kappa=0$, $\gamma_s=0$ nm$^2$ and $p=1$. Panel (b) for $\kappa=1$, $\gamma_s=20$ nm$^2$ and $p=1$. Panel (c) $\kappa=1$, $\gamma_s=20$ nm$^2$ and $p=2$. Finally, panel (d) for $\kappa=1$, $\gamma_s=20$ nm$^2$ and $p=4$. The gray region corresponds to the diamagnetic phase ($\chi<0$), whereas the white region to the paramagnetic phase ($\chi>0$).}
\label{PhaseDiagram}
\end{figure*} 
On account of the above, the total absorption coefficient is given in this order of approximation by
\bea
\alpha(\omega_0,I) =\alpha^{(1)}(\omega_0)+\alpha^{(3)}(\omega_0,I).
\label{alpha_tot_def}
\eea
The second-order nonlinear absorption coefficient vanishes whenever the medium possesses inversion symmetry, which is the case of both the parabolic and the Gaussian potentials. For the sake of further discussion, we define the critical intensity $I_c$ as the maximal value of the intensity which allows a positive defined total absorption coefficient, \textit{i.e.}, it is the defined by the conditions
\bea
\frac{\partial\alpha(\omega_0,I)}{\partial\omega_{0}}\bigg|_{\omega_{0_c},I_c}=0
, ~~~ \text{and} ~~~
\alpha(\omega_{0_c},I_c)=0.\label{Ic}
\eea
The critical intensity can be interpreted as a measure of the non-linearity of the system as it represents the point whence the one-electron density matrix can no longer be expanded in powers of the electric field strength of the optical field \cite{Ahn_Chuang}.  A lower value for $I_c$ represents an easier to obtain non-linear behavior. A large nonlinear susceptibility is indispensable in the creation of all-optical switching, modulating and computing devices \cite{Opticalproperties5, Nonlinearapps}. 
The behavior of $I_c$ as a function of the inverse kink parameter $p$ will be addressed in Section \ref{Results_discussion}.
\subsubsection{Selection Rules}
In order to find the optical properties of the system, it is necessary to compute the allowed transitions of the system. In general, one has to obtain the dipole matrix element
\bea
\langle nls|x|ml's'\rangle&=&r_0\,I_\phi(p,\Delta l)I_x(n,m,l,\Delta l,p)\,\delta_{s,s'}\nn\\
&\equiv&r_0\,\Lambda_{nm}(l,\Delta l,p)\,\delta_{s,s'},
\label{transition_def}
\eea
where we have denoted the angular momentum difference as $\Delta l=l'-l$ and $r_0=\sqrt{\hbar/m^*\Omega}$. By describing the position in the $(\rho,\phi)$ coordinate system, i.e. $x=\rho\cos\phi$, we guarantee that the electric field is only defined over the 2D-QD without introducing border effects. Subsequently, the angular integral is
\begin{subequations}
\bea
I_\phi(p,\Delta l)=\frac{1}{2\pi}\int_0^{2\pi}d\theta\,e^{i\Delta l\,\theta}\cos\alpha\theta,
\label{intphi}
\eea
while the radial integral is
\bea
&&I_x(n,m,l,\Delta l,p)=\,p\sqrt{\frac{2n!}{\left(n+p|l|\right)!}}\sqrt{\frac{2m!}{\left(m+p|l\pm\Delta l|\right)!}}\nn\\
&\times&\int dx\;e^{-x}x^{1/2}x^{\frac{p}{2}(|l|+|l\pm\Delta l|)}L_n^{p|l|}(x)L_{m}^{p|l\pm\Delta l|}(x).\nn\\
\label{Ix}
\eea
\end{subequations}
%
In Fig. \ref{Iphi} we show the result of the integral in Eq. (\ref{intphi}) as a function of $\Delta l$ for different values of the inverse kink parameter $p$.
We emphasize that the disclination introduces a nonzero probability for values of the angular momentum difference not necessarily equal to $\pm 1$, though the standard selection rule $\Delta l = \pm 1$ is recovered when no defect is introduced $(p=1)$ as the parity is well defined in such scenario. For solving the integral $I_x(n,m,l,\Delta l,p)$ from Eq.~(\ref{Ix}), we make use of the following identity~\cite{Rassias,Srivastava}:
\bea
&&\int_0^{\infty}\,dx\,e^{-x}x^\mu L_m^{\alpha}(x)L_n^{\beta}(x)=(-1)^{m+n}\Gamma(\mu+1)\nn\\
&\times&\sum_{k=0}^{\min\left\{m,n\right\}}\binom{\mu-\alpha}{m-k}\binom{\mu-\beta}{n-k}\binom{\mu+k}{k}.
\eea
%
\begin{figure*}[t]
\centering
\includegraphics[scale=0.39]{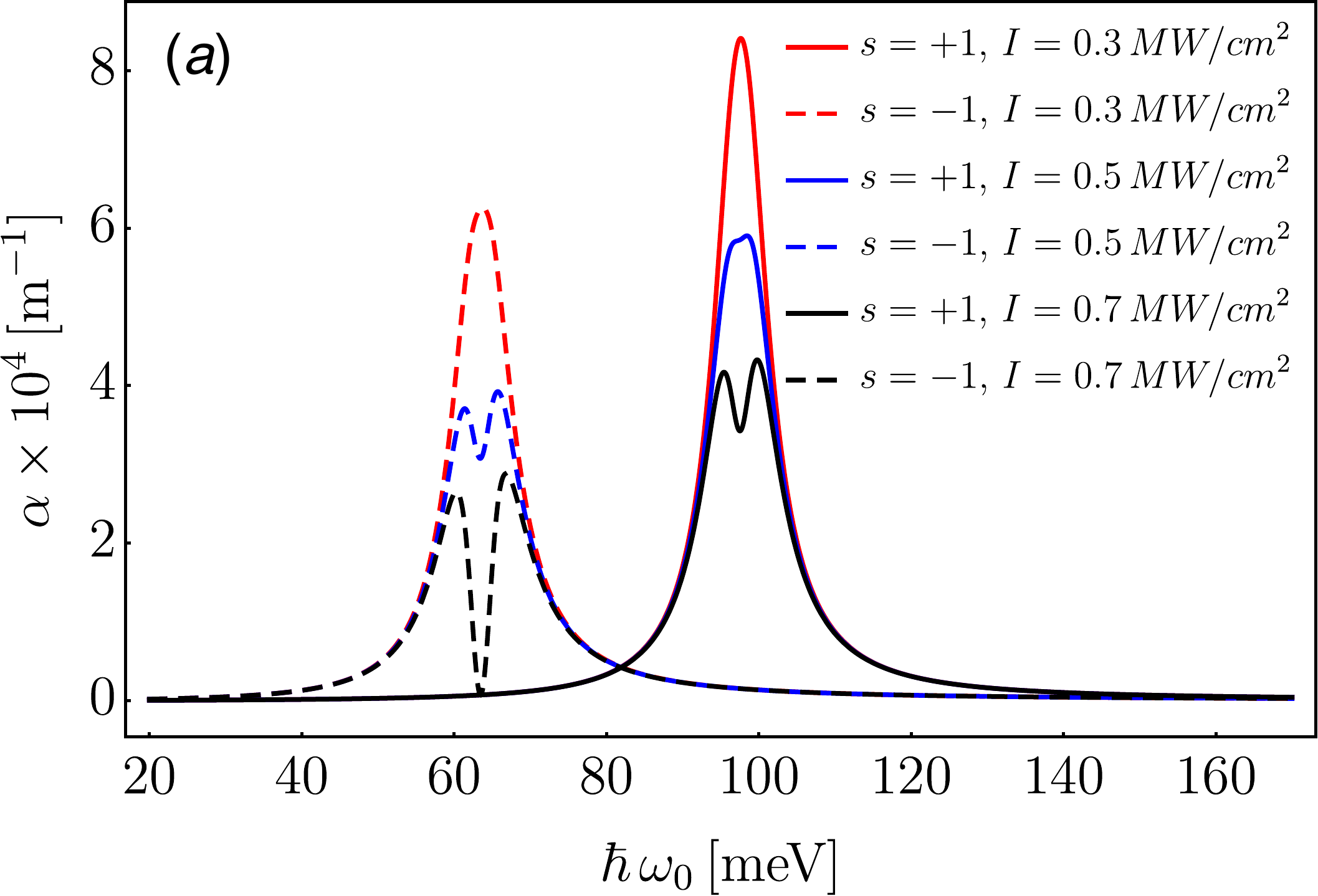}\;\;\;\;\includegraphics[scale=0.39]{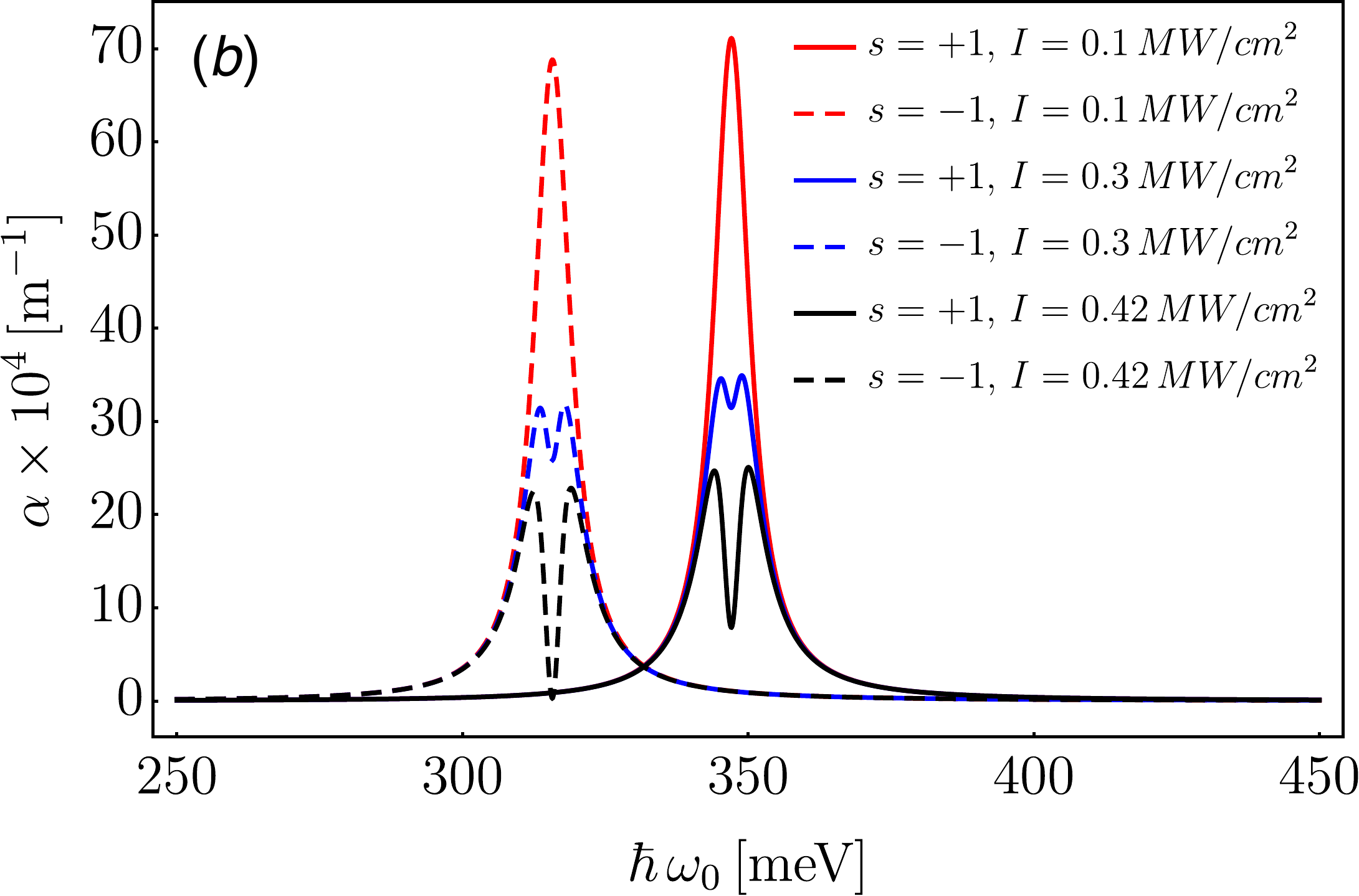}
\caption{Panel (a) shows the total absorption coefficient $\alpha$ as a function of the incident photon energy for the transition $\left|0,0,\pm1\right.\rangle\rightarrow\left|1,1,\pm1\right.\rangle$ with the inverse kink parameter fixed to $p=1$. Panel (b) shows similar numerical calculations for the inverse kink parameter fixed to $p=3$. The rest of parameters are fixed to $\Delta l=1$, $l=1$, $B=15$T, $\kappa=1$ and $\gamma_s=20$ nm$^2$. The different values of $s$ and $I$ as labelled in the legend.}\label{alpha1}
\end{figure*}
A broad set of dipolar transitions are displayed in Fig.~\ref{LambdaAB} for illustrative purposes of the long range behavior, as the topological disclination introduces the possibility of transitions with $|n-m|>1$ which were not allowed in the case $p=1$ and that are suppressed as the difference of the principal quantum numbers is increased.  The graphics show a peak in $|n-m|\approx \Delta l$ with a dominant transition of contiguous states. This implies that for $\Delta l=0$, as shown in Fig.~\ref{LambdaAB}-(a), the dominant dipole matrix elements are diagonal and thus, the system prefers to remain in the same state. In Fig. \ref{LambdaAB}-(b) it is to be noticed that the peaks get less defined and suppressed for increasing $\Delta l$. Because of the above, the number of transitions to take into account for finding the optical properties gets richer when a topological defect is introduced.
It is worth noticing that for a GaAs QD, the typical values for the Fermi energy and the level spacing is approximately to $E_{F}\approx 10$\,meV and $\Delta E \approx 20$\,meV, respectively. At low temperatures $k_B T\ll \Delta E-E_{F}\approx\,100K$ only the first sub-band lies below the Fermi energy, and in principle, there is a preferential occupation in the energetically lower states. However, theoretical studies in QDs have shown that the gain spectra and optical transitions depend strongly on temperature and the energy Fermi. It due to the combined effect of heating and carrier loss from the higher energy states, as well as from transitions from discrete levels to continuum states~\cite{Uskov:2004,Al-Khursan:2006}. Taking the above into account, our predictions about transitions from higher energy states are possible even at low temperature. We would therefore like to encourage the exploration of such transitions in our system from an experimental point of view, even for highly excited states~\cite{High_excitations}.
%
\section{Results and discussion}
\label{Results_discussion}
In order to compute the thermal and magnetic properties of our system, we use the following material constants: $m^*=0.067m_0$ the effective electron mass for a GaAs QD and 
$m_0$ being the free electron mass; $V_0=36.7$ meV is the reference potential; $g^*=-0.44$ is the effective Landé constant; $R=10$ nm is the radio of the QD; $\epsilon_r=13.18$ is the permittivity of the GaAs QD.~\cite{Opticalproperties1,Opticalproperties2,Opticalproperties3,Opticalproperties4,Opticalproperties5}.
The Rashba coupling has been taken as $\gamma_s=20$ nm$^2$ as a reasonable value in which the SOI can be appreciable~\cite{GaussianPotential1}. In order to see the behavior of the linear and third order nonlinear absorption coefficients as well as in the refractive index changes, we fix the relaxation rate to $\Gamma_{ji}=(0.2\,\text{ps})^{-1}$ as a typical value for a GaAs QD. It is worth mentioning that the relaxation time is commonly obtained from phenomenological data and it is estimated to $<1\,\text{ps}$ for a transition from the first excited state to the ground state~\cite{Zhang:2010,Liu:2012,Lu:2013b}. Moreover, due to the confinement effect in a certain direction, the electrons behave differently in the confining potential when compared to the bulk. In fact, it is well-known that in a semiconductor heterostructure the confinement effect increases with raising differences between the band gaps of the two involved materials. Furthermore, the electrons occupy discrete energy levels (discrete energy sub-bands) in the potential well along the confinement direction and behave as bulk carriers along the non-confinement directions.  Since we assume that the QD has a vertical growth direction such that $z\ll R$, the carrier density is $\sigma_v=3\times 10^{-22}$ m$^{-3}$ and remains a constant even for a varying area of the sliced sector.

The specific heat $C_v$ as a function of temperature is shown in Fig.~\ref{CvvsT}-(a) which is computed by using  Eq.~(\ref{funcionestermodinamicas}) for different values of magnetic field $B$. To appreciate the effect of both the topological disclination and the SOI interaction we compare the results of the PPM without Rasbha interaction for $p=1$ (dashed lines) with the behavior of the GPM in the presence of Rashba coupling for $p=2$ (dot-dashed lines) and $p=4$ (solid lines). In  Fig.~\ref{CvvsT}-(b) we show the behavior of $C_v$ at low temperatures, where the peak structure observed is the well-known Schottky anomaly, which occurs whenever the thermal energy gained by the electrons in the system is enough for only the lowest two levels~\cite{GaussianPotential3,SchottkyAnomalyGumber,ATari}.

From Fig.~\ref{CvvsT}-(a) it can be noticed that the specific heat converges to some finite value ($\leq 2k_B$) at high temperatures. Such limit value decreases as the magnetic field is increased and it remains even when the kink parameter is modified. Additionally, from Fig.~\ref{CvvsT}-(b) one can notice that introducing a SOI sharpens the peaks and shifts them to lower temperature values.  The reason is that the SOI lifts the spin
degeneracy and more energy levels are available in the unit range
of energy, thereby reducing the level spacing and shifting the    peak to lower thermal energy~\cite{SchottkyAnomalyGumber,ATari}.
\begin{figure}[h]
\centering
\includegraphics[scale=0.39]{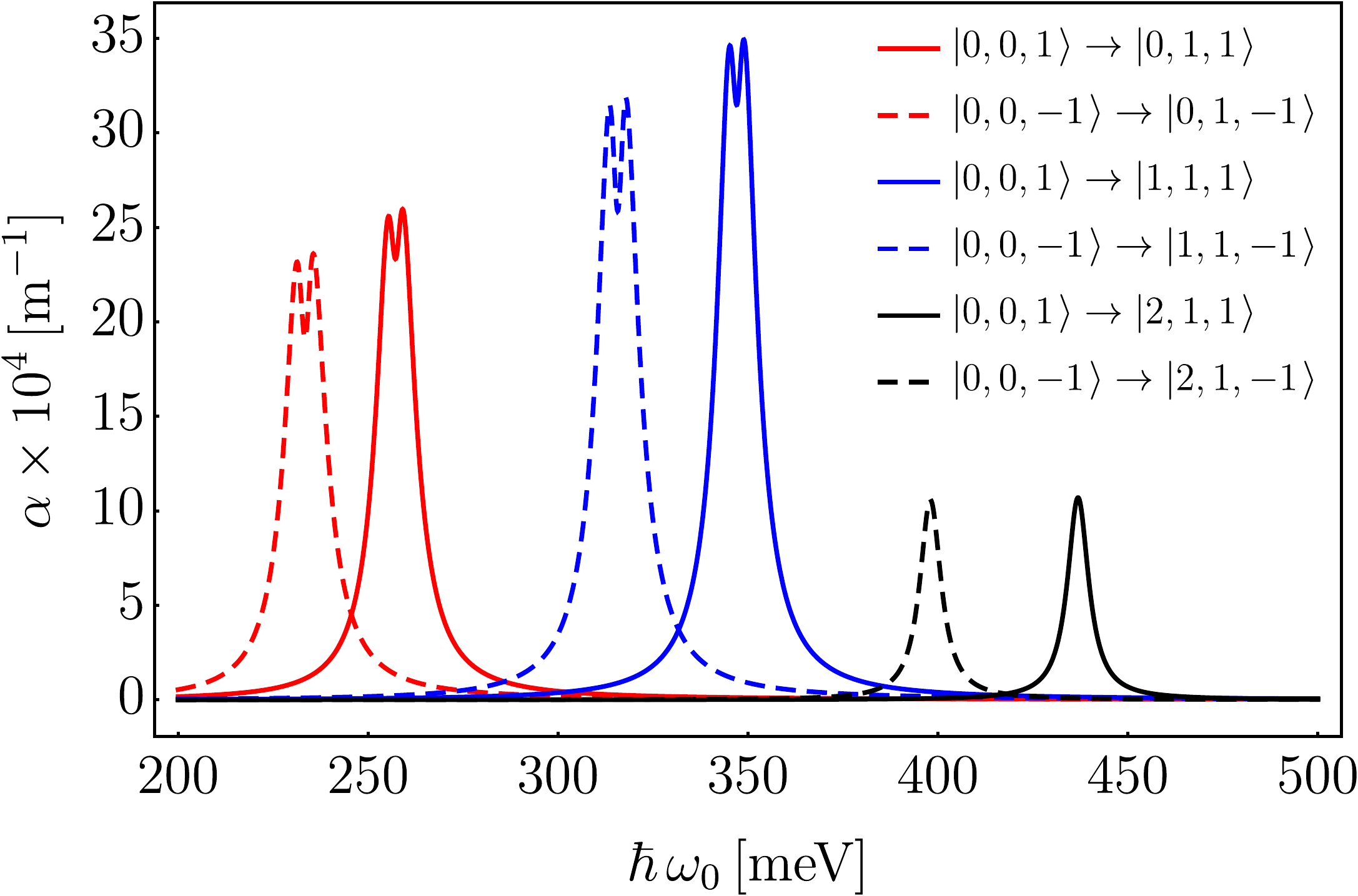}
\caption{Total absorption coefficient
$\alpha$ as a function of the incident photon energy for all allowed transitions between the lower state up to the excited state specified in the legend. The parameters are fixed to $p=3$, $I=0.3MW/cm^2$, $B=15$T, $\kappa=1$ and $\gamma_s=20$ nm$^2$.}\label{alpha2}
\end{figure}
\begin{figure*}[ht!]
\centering
\includegraphics[scale=0.39]{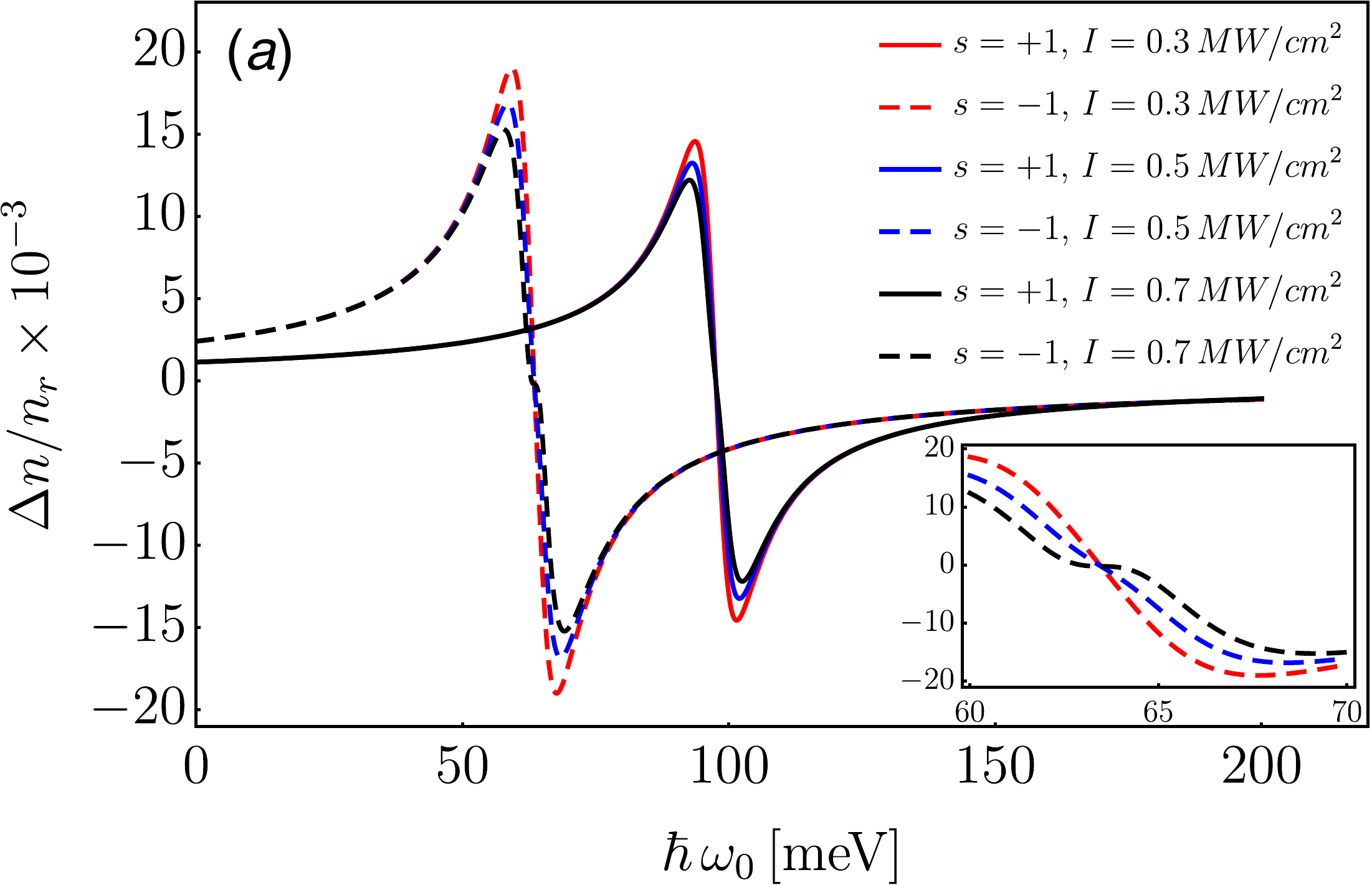}\;\;\;\;\includegraphics[scale=0.39]{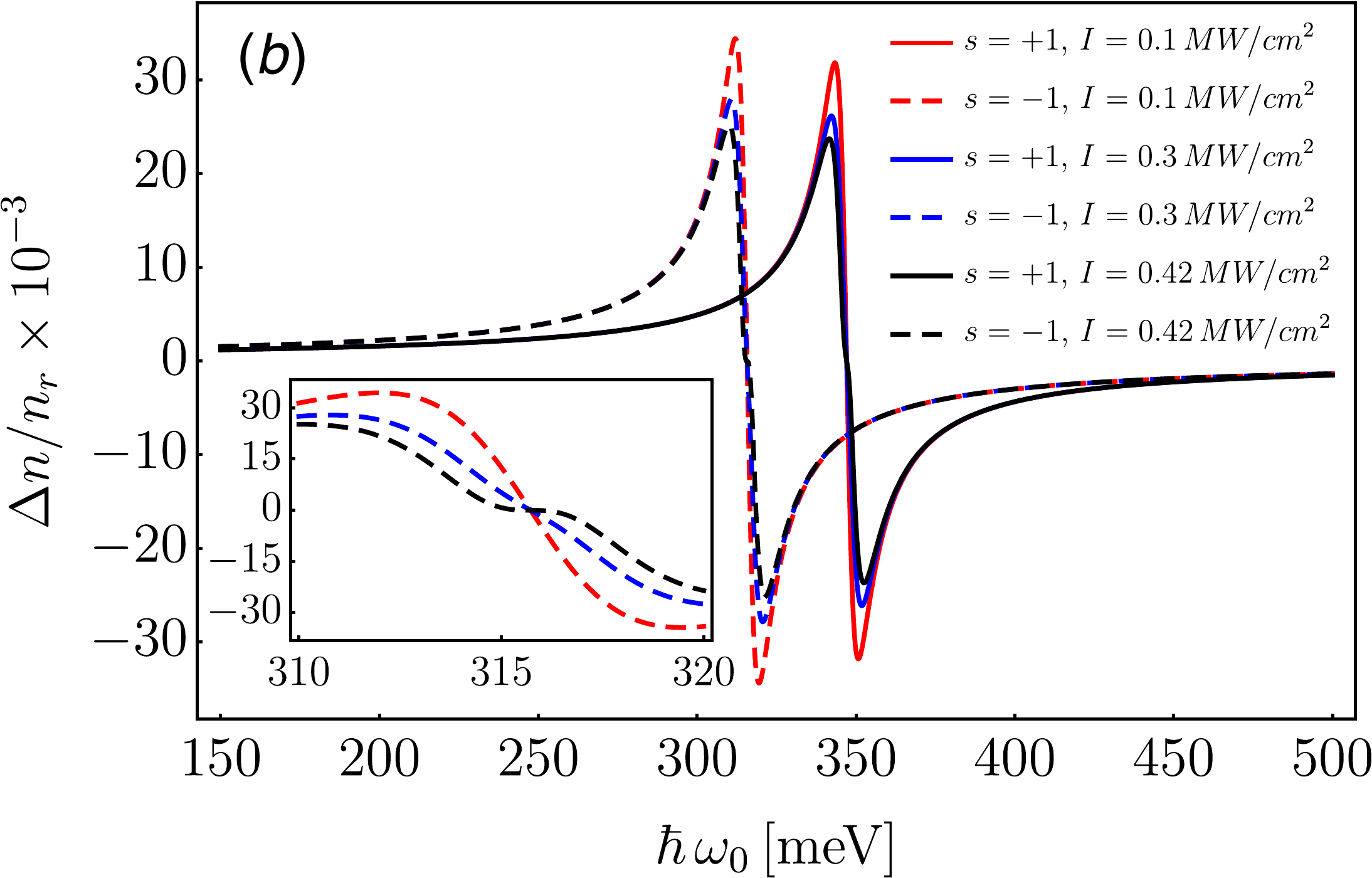}
\caption{Refraction index changes $\Delta n/n_r$ as a function of incident photon energy for the transition $\left|0,0,\pm1\right.\rangle\rightarrow\left|1,1,\pm1\right.\rangle$. The parameters were fixed to $\Delta l=1$ $l=1$, $B=15T$, $\kappa=1$ and $\gamma_s=20$ nm$^2$. Panel (a) shows the case for the inverse kink parameter $p=1$ and different values of $s$ and $I$ as labelled in the legend. Panel (b) shows similar numerical calculations for $p=3$ with different values of $s$ and $I$ as labelled in the legend. The inset diagram in panels (a) and (b) show the position of the inflection point, respectively.}\label{Deltan1}
\end{figure*}
In Fig.~\ref{PhaseDiagram} we show the thermomagnetic phase diagram obtained by evaluating Eq.~(\ref{funcionestermodinamicas}) for different values of the parameters involved. In Fig.~\ref{PhaseDiagram}-(a)
We use the PPM approximation without including the SOI and for $p=1$. Such diagram was included for comparative purposes as it represents the case without a topological disclination which has been analyzed before~\cite{Castano}. In Fig.~\ref{PhaseDiagram} (b)-(d) the graphics display the phase diagram in the GPM formalism with $\gamma_s=20$ nm$^2$ and $p=1 ,2$ y $4$, respectively. The gray region corresponds to the diamagnetic phase ($\chi<0$) and the white area to the paramagnetic phase ($\chi>0$). The effect of introducing a SOI is the decrease of the magnetic field intensity and temperature required for achieving the paramagnetic phase. Furthermore, the topological defect displaces the values of the magnetic field from the origin and gives a paramagnetic region which is not symmetric with respect to $B$. Such behavior can be understood from the energy spectrum in Eq.~(\ref{energyspectrum}): as the value of $p$ increases, the effects of the spin quantum number become less relevant and therefore, to appreciate its properties it is necessary to increase the external magnetic field or decrease the temperature of the system.
\begin{figure}[h]
\centering
\includegraphics[scale=0.39]{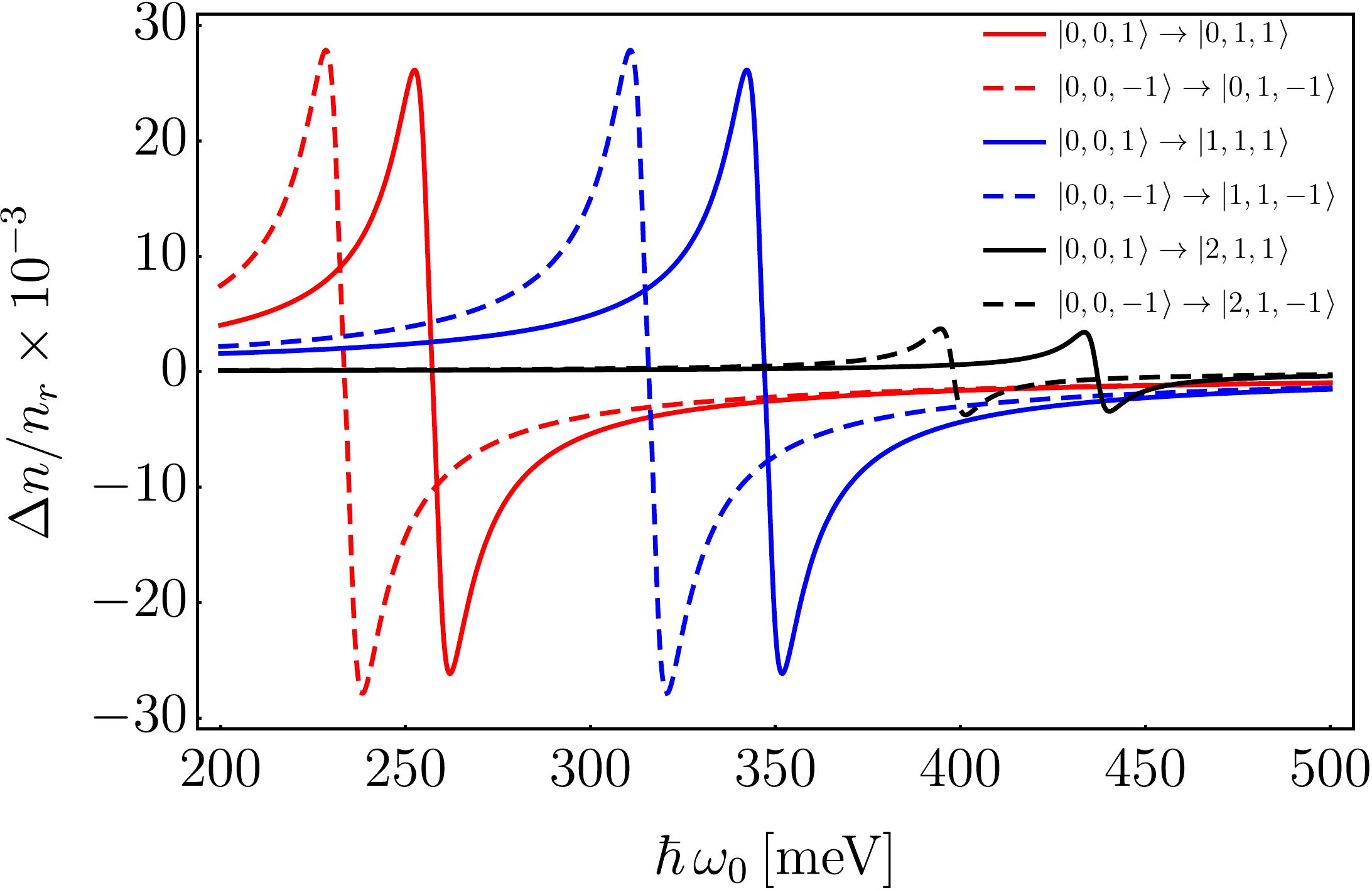}
\caption{Refraction index changes $\Delta n/n_r$ as a function of the incident photon energy for allowed transitions
between the ground state to some excited states (specified in the legend). The parameters were fixed to $I=0.3MW/cm^2$, $B=15T$, $p=3$, $\kappa=1$ and $\gamma_s=20$ nm$^2$.}\label{Deltan2}
\end{figure}
\begin{figure*}[t]
\centering
\includegraphics[scale=0.39]{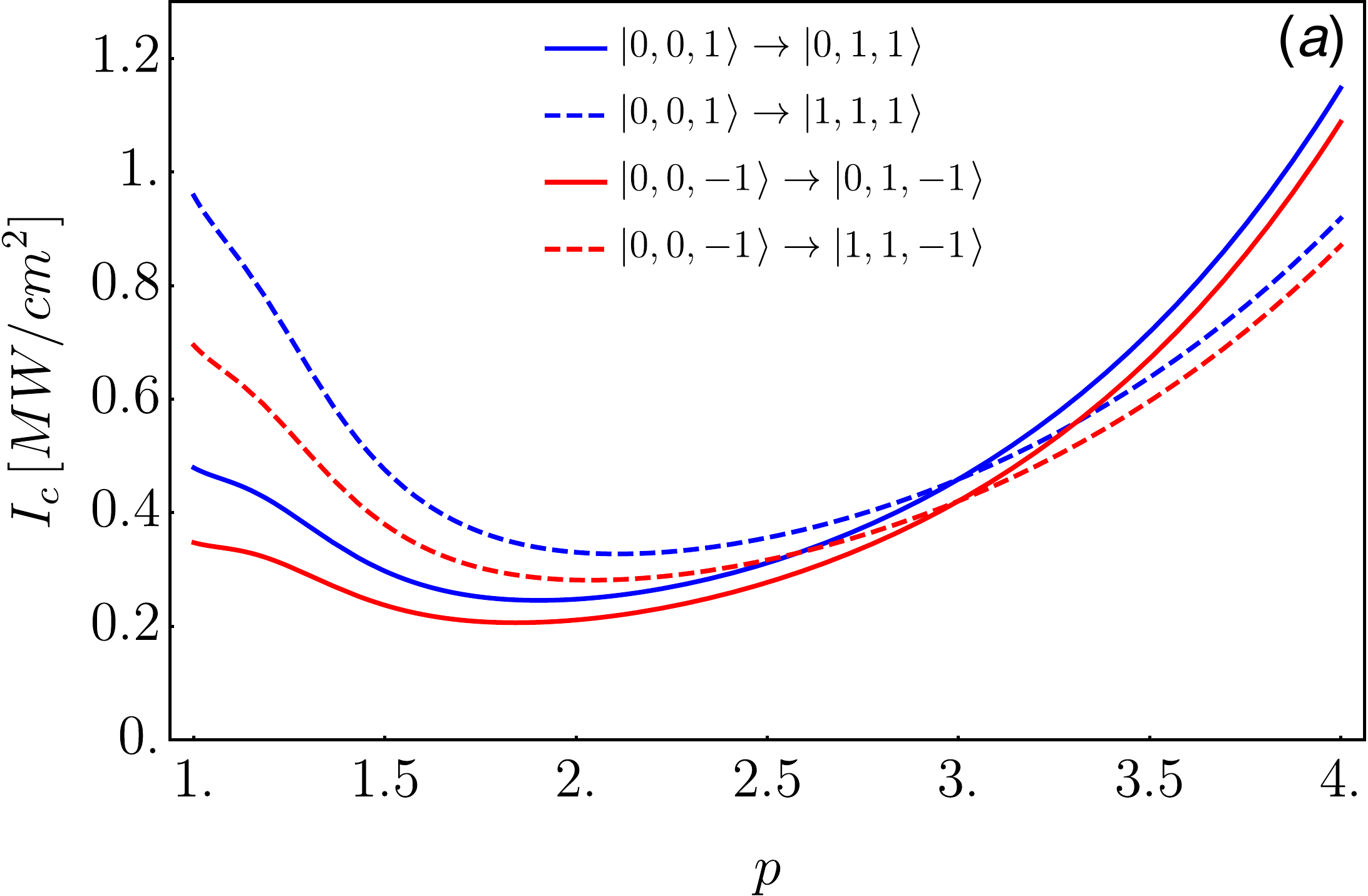}\;\;\;\;  \includegraphics[scale=0.41]{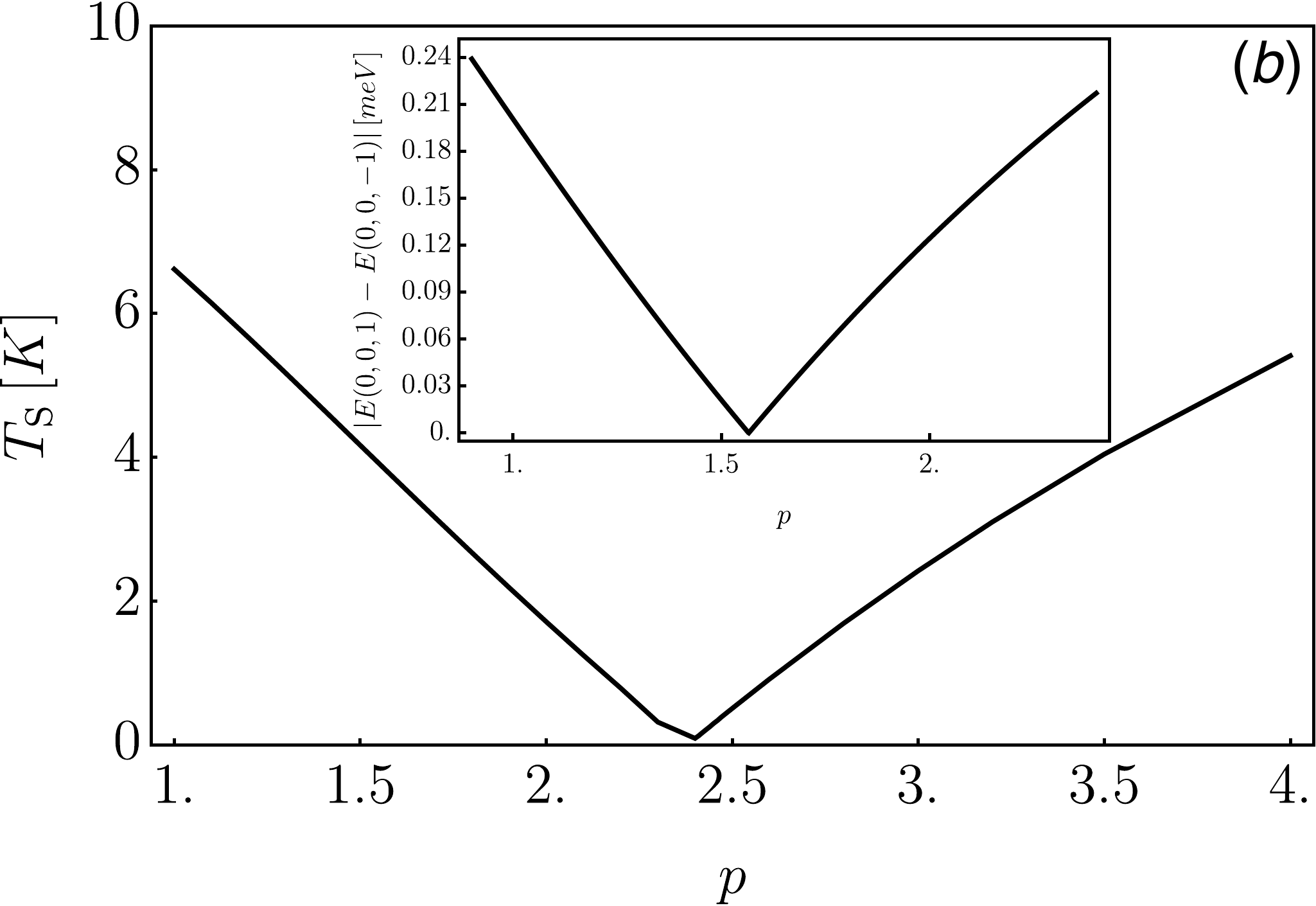}
\caption{Panel (a) shows the critical intensity $I_c$ of the incident electromagnetic wave as a function of the inverse kink parameter $p$ for different transitions (specified in the legend). Panel (b) shows the Shottky temperature as a function of $p$. The inset shows the absolute value of the energy gap between the ground and the first excited state. The parameter were fixed to $B=15 T$, $\kappa=1$ and $\gamma_s=20$ nm$^2$.}\label{IcTs}
\end{figure*}
On the other hand, Fig.~\ref{alpha1} shows the total absorption coefficient $\alpha$ as a function of the photon energy $\hbar\omega_0$ for the transition $\left|0,0,\pm1\right.\rangle\rightarrow\left|1,1,\pm1\right.\rangle$ for different values of the intensity $I$ of the external electric field.Fig.~\ref{alpha1}-(a) represents the case where no topological defect is included ($p=1$) and Fig.~\ref{alpha1}-(b) is for $p=3$. A spin dependence of the resonance can be observed, where the transitions for $s=-1$ show a peak in lower energy than the $s=1$ transitions. This effect is a consequence of the spatial wave functions depending on the spin via the definition of $\Omega_s$, as can be noticed when taking the energy difference from the spectrum in Eq.~(\ref{energyspectrum}). Such behaviour is typical in systems involving a Rashba QD in the presence of an external magnetic field~\cite{SchottkyAnomalyGumber}. In addition, the non-linear effects are enhanced by increasing the intensity of the external electric field as described by Eq.~(\ref{alpha_tot_def}). It is noteworthy that the spin-down resonances show a lower value for the critical intensity $I_c$ defined in Eq.~(\ref{Ic}). This behavior is due to $s=-1$ transitions being less energetic than those for $s=1$. Therefore, the next optical harmonic can be reached with lower external energy. 

In Fig.~\ref{alpha2} the total absorption coefficient is plotted for all the possible transitions in the case $p=3$. The same spin and intensity dependence as in Fig.~\ref{alpha1} can be observed. In addition, one can notice that the total absorption coefficient corresponding to a transition with a change in the principal quantum number $|n-m|=2$ is suppressed in magnitude and non-linear effects respect to the other transitions. 

In Fig.~\ref{Deltan1} we plot the refractive index changes $\Delta n/n_r$ as a function of the energy of the incident photon for the transition $\left|0,0,\pm1\right.\rangle\rightarrow\left|1,1,\pm1\right.\rangle$, using different values for the inverse kink parameter $p$ and the intensity of the external electric field $I$. The rest of the parameters are fixed. The index of refraction presents a blue shift when the value of the inverse kink parameter is increased. This is attributed to the confinement of the electron increasing with greater values of $p$. Besides, the characteristic energy of the spin-down states is lower than such of the spin-up states, as was previously observed in Figs.~\ref{alpha1} and \ref{alpha2}. The insets in Fig.~\ref{Deltan1}-(a) and Fig.~\ref{Deltan1}-(b) show the details of the curves with $s=-1$, where the inflection is enhanced with an increasing intensity of the external electromagnetic wave $I$ and are maximal for $I=I_c$. 

In Fig.~\ref{Deltan2} the refraction index changes are plotted for all the possible transitions in the case $p=3$. The same spin and intensity dependence as in Fig.~\ref{Deltan1} can be observed. In addition, one can notice that the $\Delta n/n_r$ corresponding to a transition with a change in the principal quantum number $|n-m|=2$ is suppressed in magnitude and non-linear effects respect to the other transitions, in concordance with the behavior shown in Fig.~\ref{alpha2}.

The Fig.~\ref{IcTs}-(a) shows the variation of the critical intensity $I_c$ defined in Eq.~(\ref{Ic}), while Fig.~\ref{IcTs}-(b) displays the Schottky temperature $T_s$ that was defined in Eq. (\ref{Schotkky_temp}). Both critical parameters are presented as a function of the inverse kink parameter $p$ and expose a change in their behavior near $p \sim 2$. We interpret this property based on the fact that $p=2$ is a special value where the wave functions have the highest symmetry. It is noticeable from Fig.~\ref{IcTs}-(a) that the non-linear effects are a non-trivial function of the inverse kink parameter $p$. Other authors have reported that such effects are caused by confinement, as this means a bigger overlap of the wave functions and therefore, bigger dipole matrix elements. Nevertheless, in the present case, the energy difference increases linearly with $p$ and thus the non-linear effects are maximal near $p=2$ before showing a decrease due to confinement \cite{Opticalproperties4, Opticalproperties5}.
The functionality shown in Fig.~\ref{IcTs}-(b) can be interpreted in terms of the lowest lying 2-level system typical for the Schottky anomaly. For a transition from the ground to the first excited state, any external  source of energy needs to be comparable with the separation of the energy levels. 
In the absence of external fields, the very first transition is given by a change of spin orientation. Because of this, the inset in Fig.~\ref{IcTs}-(b) shows the energy difference between the two lowest energy levels $E(n=0,l=0,s=\pm 1)$ obtained by evaluating Eq. (\ref{energyspectrum}) as a function of the inverse kink parameter $p$. For $p \sim 1.75$ the energy levels are close to each other and less external energy is necessary to excite the system. It is notable that the energy difference is linear in $p$, though the aspect as an absolute value function is caused by the switching of energies of the states with $s= -1$ and $s=1$. The dependence on the inverse kink parameter $p$ shown in the inset of Fig.~\ref{IcTs}-(b) does not exactly correspond to such of the main graph. The reason is that not only the thermal transition between the ground and the first excited state contribute to the Schottky anomaly, but higher energy states contribution cannot be neglected.

%
\section{Summary}\label{sec:conclusions}
In this paper we explored the effect of introducing a conical disclination on the thermal and optical properties of a two dimensional Rashba quantum dot in the presence of a uniform and constant magnetic field. The obtained specific heat presents the Schottky anomaly in the low temperature regimen while the Schottky peaks are displaced linearly respect to the inverse kink parameter, which controls the topological defect. Such defect and the Rashba coupling modify the values of temperature and magnetic field in which the system behaves as a paramagnetic material.
Furthermore, the introduction of a deficit angle relaxes the selection rules for the dipolar transitions, as the parity of the eigenfunctions is mixed. This unusual behavior can be observed in the absorption coefficient and the refraction index changes as semi-suppressed blue-shifted resonances. The spin dependence of the optical-properties caused by the Rashba SOI and the enrichment of the allowed transitions may open a set of technological applications. Finally, we introduced the concept of critical intensity as a measure of non-linear effects and showed that the topological defect enhances them approximately when half of the QD is sliced off.\\

\section{ACKNOWLEDGMENT}
The authors J. D. Castaño-Yepes, D. A. Amor-Quiroz and C. F. Ramirez-Gutierrez acknowledge financial support from Consejo Nacional de Ciencia y Tecnolog\'ia (CONACyT). E.A.G acknowledges financial support through the project No. 919 from the Vicerrector\'ia de Investigaciones at the Universidad del Quind\'io. The authors also thank Dr. Rafael Quintero Torres and Dr. Cedric Lorcé for useful comments about the optical properties.
\\\\

\end{document}